\documentclass[11pt,a4paper,twoside]{article}
\usepackage[T1]{fontenc}
\usepackage[utf8]{inputenc}
\usepackage{amssymb}
\usepackage{lipsum}
\usepackage{authblk}
\usepackage[top=3cm, bottom=3cm, left=2cm, right=2cm]{geometry}
\usepackage{fancyhdr}
\usepackage{adjustbox}
\usepackage[sort&compress, numbers, compress ]{natbib}
\usepackage[colorlinks,bookmarksopen,bookmarksnumbered,citecolor=red,urlcolor=red]{hyperref}
\usepackage{algorithm}
\usepackage{algorithmic}
\usepackage{boxedminipage}
\usepackage{comment}
\usepackage{eqparbox}
\usepackage{multirow}
\usepackage{multicol}
\usepackage{caption}
\usepackage{subcaption}

\usepackage{nomencl}
\makenomenclature

\usepackage{amsmath}
\usepackage{amsfonts}

\DeclareMathOperator*{\argmin}{arg\,min}

\usepackage{graphicx}

\usepackage{tikz}

\newlength{\commentindent}
\usepackage{array}
\newcolumntype{R}[1]{>{\raggedleft\let\newline\\\arraybackslash\hspace{0pt}}m{#1}}

\newcommand\BibTeX{{\rmfamily B\kern-.05em \textsc{i\kern-.025em b}\kern-.08em
T\kern-.1667em\lower.7ex\hbox{E}\kern-.125emX}}

\newcommand{\dive}{\nabla \cdot}
\newcommand{\grad}{\nabla}

\newcommand{\EL}{\mathcal{L}}
\newcommand{\bx}{\mathbf{x}}
\newcommand{\bu}{\mathbf{u}}

\let\OLDthebibliography\thebibliography
\renewcommand\thebibliography[1]{
	\OLDthebibliography{#1}
	\setlength{\parskip}{0pt}
	\setlength{\itemsep}{0pt plus 0.3ex}
}

\renewenvironment{abstract}{%
	\begin{center}
		\begin{adjustbox}{minipage=0.90\textwidth,center} 
			\rule{\textwidth}{0.5pt}}
		{\par\noindent\rule{\textwidth}{0.5pt}  
		\end{adjustbox}
	\end{center} 
}

\makeatletter

\renewcommand\@maketitle{%
	\begin{adjustbox}{minipage=0.90\textwidth,center}
		\begin{center}
			\vskip 1em
			\let\footnote\thanks 
			{\LARGE \@title \par }
			\vskip 1.0em
			{\large \@author \par}
		\end{center}
	\end{adjustbox}
	\vskip 0.0em \par
}
\makeatother
\usepackage{etoolbox}
\renewcommand\nomgroup[1]{%
  \item[\bfseries
  \ifstrequal{#1}{G}{Greek Symbols}{%
  \ifstrequal{#1}{S}{Symbols}{%
  \ifstrequal{#1}{A}{Abbreviations}{%
  \ifstrequal{#1}{B}{Subscripts}{%
  \ifstrequal{#1}{P}{Superscripts}{
  }}}}}]}

\begin{document}

\title{Physics-Informed Neural Networks for Mesh Deformation with Exact Boundary Enforcement}
\author[1]{A.~Aygun\footnote{Corresponding author. E-mail address: atakana@metu.edu.tr} }
\author[2]{R.~Maulik}
\author[1]{A.~Karakus}

\affil[1]{\textit{\small{Department of Mechanical Engineering, Middle East Technical University, Ankara, Turkey 06800}}}
\affil[2]{\textit{\small{Mathematics and Computer Science Division, Argonne National Laboratory, Lemont, IL, 60439, USA}}}

\renewcommand\Authands{ and }
\date{\vspace{-5ex}}
\maketitle

\begin{abstract}
In this work, we have applied physics-informed neural networks (PINN) for solving mesh deformation problems. We used the collocation PINN method to capture the new positions of the vertex nodes while preserving the connectivity information. We use linear elasticity equations for mesh deformation. To prevent vertex collisions or edge overlap, the mesh movement in this work is conducted in steps with relatively small movements. For moving boundary problems, the exact position of the boundary is essential for having an accurate solution. However, PINNs are frequently unable to satisfy Dirichlet boundary conditions exactly. To overcome this issue, we have used hard boundary condition enforcement to automatically satisfy Dirichlet boundary conditions. Specifically, we first trained a PINN with soft boundary conditions to obtain a particular solution. Then, this solution was tuned with exact boundary positions and a proper distance function by using a new PINN considering only the equation residual. To assess the accuracy of our approach, we used the classical translation and rotation tests and compared them with a proper mesh quality metric considering the change in the element area and shape. The results show the accuracy of this approach is comparable with that of finite element solutions. We also solved different moving boundary problems, resembling commonly used fluid-structure interaction problems. This work provides insight into using PINN for mesh-deformation problems without needing a discretization scheme with reasonable accuracy. 

\end{abstract}


\section{Introduction}
Dynamic grids in numerical fluid flow simulations generally arise in many applications, such as airfoil movement \cite{batina1990dynamicAirfoil, robinson1991dynamicAirfoil}, blood flow \cite{bazilevs2006bloodFsi}, parachute mechanics \cite{stein2000parachute, tezduyar2008parachute}, and free surface flow problems \cite{tezduyar1992freesurface}. These and other fluid-structure interaction (FSI) problems need to move the computational grid with moving boundaries. The naive choice is to regenerate the mesh every time the boundary moves. Regenerating the mesh for a complex geometry results in a need for an automatic mesh generator \cite{tezduyar2001movingFEM}. This approach alters the grid connectivity and, therefore, brings up a need to project the solution to the new mesh. This introduces new projection errors each time the mesh is updated. Moreover, the cost of calling a new mesh generation algorithm can be overwhelming, especially for 3D problems \cite{johnson1994meshUpdate}.

Specific mesh moving techniques can overcome the drawbacks of remeshing for moving boundary problems. These methods try to update the position of the nodes of the original mesh under some prescribed laws without changing the grid connectivity. Farhat et al. introduced a spring analogy, where they fictitiously attach a torsional spring to the nodes of the mesh \cite{farhat_torsional_1998}. The system has fictitious mass, damping, and stiffness matrices, and the forcing is the displacement of the moving boundaries. This approach prevents vertex collisions as well as penetrating grid edges. In \cite{johnson1994meshUpdate}, the authors used a linear elastic equation to represent the fluid domain as an elastically deformable body and introduced a parallel finite element strategy. Using the same elasticity formulation, Stein et al. \cite{stein2003lineMesh} solved the equation using a Jacobian-based stiffening. They introduced an additional stiffening power as a function of transformation Jacobian in the finite element formulation. This addition allowed them to stiffen the smaller elements more than the larger ones, resulting in improved mesh quality near the moving surfaces. Takizawa et al. \cite{takizawa_low-distortion_2020}, introduced a method based on the fiber-reinforced hyperelasticity model. They introduced fibers in different directions according to the motion, which allows the model to reduce the distortion of a mesh element. The moving mesh problem can be solved using the Laplacian or biharmonic equations \cite{lohner1996laplacian, robertson1999laplacian, helenbrook2003biharmonic}. Although using the biharmonic operator introduces extra computational complexity compared to the Laplacian equation systems, it can give the extra ability to control the normal mesh spacing \cite{helenbrook2003biharmonic}.

Apart from conventional numerical methods, machine learning methods are also used to solve partial differential equations. Deep neural networks were first used by Lee and Kang \cite{lee1990neuralAlgo}, and Lagaris et al \cite{lagaris1998artificial} to predict the solution of a partial differential equation (PDE). Raissi et al. \cite{raissi_pinn_2019} introduced the concept of physics-informed neural networks (PINN) to solve PDEs without any given data. This approach gives information about the physical laws to the neural network. Using the information on the boundary and initial conditions, neural networks can predict the solution of a PDE. The PINN formulation has received great attention and has been studied in wide content. There are numerous extensions of PINN to improve the methodology. Several domain decomposition models are designed to improve the accuracy and allow parallelization \cite{karniadakis_extended_2020, kharazmi_hp-vpinns_2021, shukla_parallel_2021}. Bayesian PINNs are proposed to tackle the problems involving solving PDEs where noisy data is available \cite{yang_b-pinns_2021} and where uncertainty quantification is important.  Applications of PINN cover the solutions of conservation laws \cite{jagtap2020conservativePinn}, fractional and stochastic differential equations \cite{pang2019fpinns, yang2020stochasticPinn}, solution of Navier-Stokes equations \cite{raissi2020hidden, jin2021nsfnets}, Euler equations \cite{mao2020highSpeed}, heat transfer problems \cite{cai2021pinnHeat, aygun2022pinn}, Boltzmann equation with Bhatnagar-Gross-Krook collision model \cite{lou2021bgkPinn}, Allen-Cahn and Cahn-Hilliard equations \cite{wight2020pinnCahn, mattey2022pinnCahn}, free boundary and Stefan problems \cite{wang2021deepStefan} and many more. 

Despite the success of PINN across a range of different problems, it can face difficulties when solving multiscale and multiphysics problems \cite{karniadakis2021pinnReview}, especially for dynamical systems with chaotic or turbulent behavior \cite{wang2022respecting}. The fully connected networks face difficulties in learning high-frequency functions. This phenomenon is named spectral bias \cite{rahaman_spectral_2019, wang2022ntk_pinn}. The high-frequency behavior in the objective function results in sharp gradients. Therefore, PINN models can have difficulties while penalizing the residual loss. Although there are several approaches to tackle these problems and improve the training capabilities of PINN, the classical PINN method shows better performance to accurately solve the PDEs that govern the mesh deformation. Therefore our research focuses on using PINNs in the application of these problems.

The main objective of this paper is to show the applicability of physics-informed neural networks for moving mesh problems. The PINN approach can produce satisfactory solutions for the movement of boundaries without needing a discretization scheme. However, using the original PINN formulation for mesh moving problems can have difficulties with the static and moving boundaries. PINN minimizes the loss at the boundaries, without imposing boundary conditions exactly. To overcome this problem, we used exact boundary enforcement. After obtaining a particular solution that weakly satisfies the boundary conditions, the prediction is corrected by training another PINN. To the best of our knowledge, using PINNs on mesh movement problems with exact boundary enforcement is not studied in detail in the literature.

The remainder of this paper is organized as follows. First, basic information is given about physics-informed neural networks. This chapter is enhanced with the methodology of automatically satisfying boundary conditions using exact boundary enforcement. Most common mesh movement techniques are presented in the next chapter alongside the mesh quality metric used for comparing different methods. The results are presented with classical translation and rotation tests followed by examples resembling commonly used moving boundary problems.

\section{Physics-Informed Neural Networks}
A basic, fully connected deep neural network architecture can be used to solve differential equations \cite{lagaris1998nnPde}. Given an input vector $\bx \in \mathbb{R}^d$, a single layer neural network gives an output $\hat{\bu}$ by the following form:

\begin{equation}
    \label{eq:singleLayerNN}
    \hat{\bu} = \sigma(\mathbf{W}_1\bx + \mathbf{b}_1)\mathbf{W}_2 + \mathbf{b}_2,
\end{equation}
where $\mathbf{W}$ are the weight matrices and $\mathbf{b}$ are the bias vectors. $\sigma(\cdot)$ is a nonlinear function known as the activation function. In general, Sigmoid, hyperbolic tangent, and rectified linear unit (ReLU) are popular choices for the activation function. The hyperparameters $\theta = [\mathbf{W}, \mathbf{b}]$ are estimated by the following optimization problem

\begin{equation}
    \label{eq:optimization}
    \mathbf{\theta}^* = \argmin_\mathbf{\theta} J(\mathbf{\theta}; \mathbf{x}).
\end{equation}
Here, $J$ is the objective function to be minimized. In this work, this function is defined as the mean squared error of the prediction. This minimization problem in Equation \ref{eq:optimization} can be solved by using first-order stochastic gradient descent (SGD) algorithms \cite{ruder2016sgd}. In each iteration, the hyperparameters are updated in such a way, 

\begin{equation}
    \label{eq:SGD}
    \theta^{i+1} = \theta^i - \eta^i \nabla_{\theta}J(\theta; \bx),
\end{equation}
where $i$ being the current iteration and $\eta$ is the learning rate. The gradient of the loss function, $\nabla_{\theta}J(\theta; \bx)$, is calculated by backpropagation \cite{rumelhart1986backprop}. 

For the physics-informed a neural network, we consider the general form of partial differential equations:
\begin{subequations}
\label{eq:PDEsystem}
\begin{align}
 &\mathbf{u}_t + \mathcal{N}[u] = 0, \quad &\mathbf{x} \in \Omega,\; t\in[0,T]\label{eq:PDE}\\
 &\mathbf{u}(\mathbf{x},0) = f(\mathbf{x}), \quad &\mathbf{x} \in \Omega\label{eq:PDE IC}\\
 &\mathbf{u}(\mathbf{x},t) = g(\mathbf{x},t), \quad &\mathbf{x} \in \partial\Omega, \; t \in [0,T]\label{eq:PDE BC}
\end{align}
\end{subequations}
where $\mathcal{N}$ is a generalized differential operator that can be linear or nonlinear, $\bx \in \mathbb{R}^d$ and $t \in [0, T]$ are the spatial and temporal coordinates. $\Omega \;\text{and}\; \partial\Omega$ represent the computational domain and the boundary respectively. $\mathbf{u}(\bx,t)$ is the general solution of the PDE with $f(\bx)$ is the initial condition and $g(\bx,t)$ is the boundary condition. The hidden solution, $\bu(\bx, t)$, can be approximated under the PINN framework proposed by Raissi et al. \cite{raissi_pinn_2019}, by a feedforward neural network $\hat{\bu}(\bx, t; \theta)$ with parameters $\mathbf{\theta}$. For the supervised training the only labeled data comes from the boundary/initial points. Inside the domain, the loss is determined by the PDE residual. By utilizing automatic differentiation (AD) \cite{baydin_automatic_2017}, PINNs can differentiate the network output w.r.t the input layer. AD applies the chain rule repeatedly to the elementary functions and arithmetic operations to achieve
the derivative of the overall composition. AD is well implemented in popular deep learning frameworks such as TensorFlow \cite{abadi2016tensorflow} and PyTorch \cite{paszke2019pytorch}.\

\begin{figure}[htbp!]
    \centering
    \includegraphics[width=\textwidth]{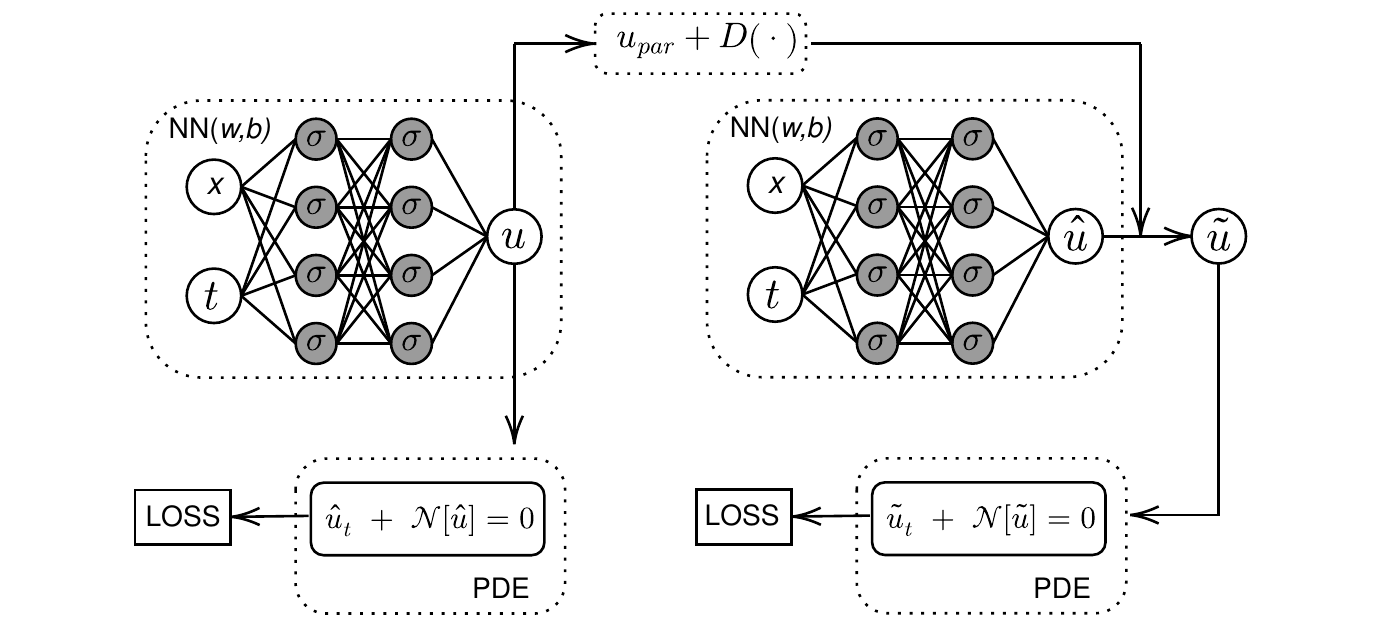}
    \caption{Schematic of PINN approach with exact boundary enforcement. The first PINN on the left shows the original formulation with weakly enforced Dirichlet boundary conditions. The second network uses the particular solution with exact boundary enforcement to satisfy Dirichlet boundaries exactly}
    \label{fig:PINN}
\end{figure}

In classical PINN implementations, the loss term is a composite term including supervised data loss on the boundaries and initial points and the PDE loss. The total loss term can be written such that, 
\begin{equation}
\label{eq:pinnLoss}
    \EL = w_{R}\EL_{R} + w_{BC}\EL_{BC} + w_{IC}\EL_{IC}.
\end{equation}
Here the terms represent the boundary loss $\EL_{BC}$, the initial condition loss $\EL_{IC}$, and the PDE residual loss $\EL_{R}$. The $w$ terms are specific weights of each loss term that can be user-specified or tuned manually or automatically \cite{wang2021_gradientflow, wang2022ntk_pinn}. Each loss term can be written as, 
\begin{subequations}
\label{eq:LossFunctions}
\begin{align}
 \mathcal{L}_R &= \frac{1}{N_R}\sum_{i=1}^{N_R}|\mathbf{u}_t + \mathcal{N}[\mathbf{u}(\mathbf{x}^i,t^i)]|^2\\
 \mathcal{L}_{BC} &= \frac{1}{N_{BC}}\sum_{i=1}^{N_{BC}}|\mathbf{u}(\mathbf{x}^i,t^i) - g(\mathbf{x}^i,t^i)|^2\\
 \mathcal{L}_{IC} &= \frac{1}{N_{IC}}\sum_{i=1}^{I_{BC}}|\mathbf{u}(\mathbf{x}^i,0) - f(\mathbf{x}^i)|^2.
\end{align}
\end{subequations}
Here $N_R, \; N_{BC}$, and $N_{IC}$ are the total number of points used for calculating the mean squared error used here as the loss function. The schematic of a classical PINN can be seen in the left part of Figure \ref{fig:PINN}.

\noindent The hyperparameters $\mathbf{\theta}=[\mathbf{W}, \mathbf{b}]$ can be optimized by a chosen optimization algorithm to find the minimum total loss defined in Equation \ref{eq:pinnLoss}. As mentioned above, stochastic gradient descent algorithms are commonly used in neural network implementations \cite{ruder2016sgd}. This method aims to find new parameters $\theta$ in the opposite direction of the gradient of the objective function. The gradient of the loss function w.r.t. hyperparameters is calculated by backpropagation. In this work, we used the ADAM algorithm \cite{kingma2014adam} as the SGD optimizer. 

\subsection{Exact Boundary Enforcement}


The optimization algorithm used in PINN tries to minimize the physics-based loss, $\EL_R$. Using proper boundary and initial conditions can regularize the physics loss in deep neural networks. This classical PINN boundary condition implementation in Equation \ref{eq:LossFunctions} is named soft boundary enforcement \cite{sun2020hardbcPinn}. In this approach, the boundary prediction is minimized in the composite loss function. Although the SGD algorithms can minimize these loss functions, they do not satisfy the boundary values exactly. However, some PDE applications, such as mesh movement, need exact boundary values. For this purpose, we apply exact boundary enforcement. Sun et al. \cite{sun2020hardbcPinn} used this boundary condition enforcement to exactly satisfy the velocity and pressure values on the boundaries of internal flow cases with Navier-Stokes equations. Sukumar and Srivastava \cite{sukumar2022exact} introduced geometry-aware trial functions. They multiply the neural network output with these functions and use its generalization to exactly satisfy boundary conditions on complex geometries. In this work, we use this idea with multiple physics-informed neural networks to exactly satisfy Dirichlet boundary conditions. First, we trained a PINN with soft boundaries. For the mesh movement problem, the displacement vector ${\bu} = [X, Y]^T$ will give the new coordinates of the nodes from the first neural network prediction. This solution is then changed on the boundaries with the exact values. This new solution is the particular solution of our approach. Then, a new PINN is trained with an output $\hat{\bu}(\bx;\theta)$. This output is modified with the following equation.
\begin{equation}
    \label{eq:hardBC}
    \tilde{\bu}(\bx; \theta) = \bu_{par}(\bx) + D(\bx)\hat{\bu}(\bx;\theta).
\end{equation}
Here, $\bu_{par}$ is a particular solution that is a globally defined smooth function that only satisfies the boundary conditions. Any smooth function can be used for the particular solution such as radial basis functions (RBF) or linear functions \cite{sun2020hardbcPinn}. In this work, we use the classical PINN predictions with the soft boundary condition implementation as the particular solution. $D$ is a specified distance function from the boundary. Equation \ref{eq:hardBC} states that on the boundaries $D(\bx)=0$, the particular solution satisfies the exact boundary values, $\bu = g$ on $\partial\Omega$. For a general approach, we used the shortest distance between the residual points and the boundaries. Since the geometric domains used in this paper are not too complex, this approach is not very time consuming. For complex geometries, approximate distance functions using R-functions \cite{sukumar2022exact} or pre-trained deep neural networks \cite{berg2018unified} can be used. This modified output contributes to the physics loss of the new PINN. In this network, the objective function is only consisting of the PDE residual $\EL_R$ and trained with the same PDE. This approach allows us to exactly satisfy the Dirichlet boundary conditions using PINN.

\section{Mesh Movement}
\label{ch:meshMotion}
Mesh movement strategies to deform the mesh with a moving boundary generally can be performed by solving a PDE or using an interpolation scheme \cite{luke_fast_2012}. All of these techniques have the goal to provide a displacement of the moving boundary and propagate this movement into the domain. Methods with a PDE solution, generally model the mesh movement as a physical process which can be solved using numerical methods. One of the popular versions includes modeling the domain with torsional springs that prevent the vertices to collide \cite{farhat_torsional_1998}. In a similar manner, this movement can be modeled with an elastic \cite{johnson1994meshUpdate, stein2003lineMesh} and hyperelastic \cite{takizawa_low-distortion_2020} analogy, where the computational domain is simulated as an elastic body. Nonlinear elasticity equations with neo-Hookean models can be used in the same way as the elastic equations \cite{shamanskiy2021mesh}. Other techniques include mesh deformation as a diffusive system modeled with the Laplacian or biharmonic equations \cite{helenbrook2003biharmonic}. All these PDEs can be solved using traditional numerical methods such as FEM. 

Interpolation schemes consider the mesh movement as a problem of interpolation from the boundaries to the domain. These schemes use interpolation on scattered data and generally do not need connectivity information. Using radial basis functions (RBF) is one of the common methods. In \cite{de_boer_mesh_2007}, de Boer et al. use RBF interpolation on unstructured grids to estimate the movement. The equation system only involves the boundary nodes and displacement of the whole mesh is modeled. Extending this method, in \cite{rendall_efficient_2009}, the authors use data reduction algorithms using a coarse subset of the surface mesh. With greedy algorithms, this approach is effective, especially for mesh motion problems with smooth surface deformations. 

In this work, we used one of the common PDEs for mesh movement. The mesh motion is calculated by using the linear elasticity equation from structural mechanics. The coordinates of the nodes will be defined as $\bu$, the computational domain is referred to as $\Omega$, and the boundaries are $\partial\Omega$. Boundaries also include the moving objects inside the meshes. The new coordinates of the moving and stationary boundaries are given as the Dirichlet boundary condition. The movement of an object inside the mesh deforms the computational domain which is modeled as an elastic body. The new coordinates can be found by the following linear elasticity equation:

\begin{subequations}
\label{eq:LinearElasticity}
\begin{align}
    \dive \boldsymbol{\sigma}(\bu) = 0 \quad &\mathrm{in}\; \Omega\\
    \bu = \bu_b \quad &\mathrm{on}\; \partial\Omega.
\end{align}    
\end{subequations}
Here $\boldsymbol{\sigma}$ is the Cauchy stress tensor. It is related to the strain tensor $\boldsymbol{\epsilon} = (\grad\bu + \grad\bu^T)/2$. The stress tensor can be written in a way by Hooke's law:
\begin{equation}
    \label{eq:CauchyTensor}
    \boldsymbol{\sigma} = \lambda \mathrm{tr}(\boldsymbol{\epsilon})\mathbf{I} + 2\mu\boldsymbol{\epsilon}.
\end{equation}
The Lam\'e parameters $\lambda$ and $\mu$ are structural parameters coming from the elastic modulus $E$ and Poisson's ratio $\nu$. Since the mesh domain is not a real elastic body, the exact values for these parameters are not known. A value between 0.3 and 0.45 is recommended for Poisson's ratio since a high value can lead to distorted elements, and a lower value can reduce the resistance \cite{shamanskiy2021mesh}. 

To be able to compare the effectiveness of different mesh movement techniques after a deformation, we use a mesh quality metric based on \cite{stein2003lineMesh}. In these metrics, the area and shape changes are considered by checking the element area and the aspect ratio. Both metric uses the initial mesh elements as reference elements and measures the change according to them. The element area change $f_A^e$ and shape change $f_{AR}^e$ is defined as :
\begin{subequations}
    \label{eq:meshMetric}
    \begin{align}
        f_A^e &= \left\lvert \log \left(\frac{A^e}{A_o^e}\right) / \log (2.0) \right\rvert, \\
        f_{AR}^e &= \left\lvert \log \left(\frac{AR^e}{AR_o^e}\right) / \log (2.0) \right\rvert. 
    \end{align}
\end{subequations}
Here, the superscript $e$ represents the specific element, and the subscript $o$ is the initial mesh element before the deformation occurs. $AR^e$ is the element aspect ratio defined in \cite{stein2003lineMesh} as:
\begin{equation}
    AR^e = \frac{(l^e_{max})^2}{A^e}.
\end{equation}
Here, $l^e_{max}$ is the maximum edge length for the specific element. For comparison of different techniques, we use the global area and shape changes by considering the maximum values of element area and shape changes, respectively.

\section{Results}
The movement of dynamic meshes with PINN is presented with several different test cases. First, a deformed square is presented where we squeeze the domain from the top and bottom. Then, the basic translation and rotation tests are performed and the solutions of the PINN approach are compared with the finite element solutions. Lastly, the movement of a flexible beam is presented where one end of the beam is fixed. For all the problems, initial meshes are generated using the Gmsh mesh generator \cite{geuzaine2009gmsh}. We used TensorFlow to construct our PINN framework  with Adam optimizer as the gradient descent algorithm. We initialized all the neural networks using the Glorot scheme and used 7 hidden layers with 50 units. The classical neural networks are trained for 40000 iterations, and the networks with exact boundary enforcement are trained for 5000 iterations. The learning rate is $10^{-3}$ with a decay rate of 0.9. The Lam\'e parameters are selected as $\mu=0.35$ and $\lambda=1$ as recommended \cite{shamanskiy2021mesh}.

\subsection{Deformed Square}
\begin{figure}[hbtp!]
	\begin{center}
    \begin{subfigure}[b]{0.32\textwidth}
	\includegraphics[width=\textwidth]{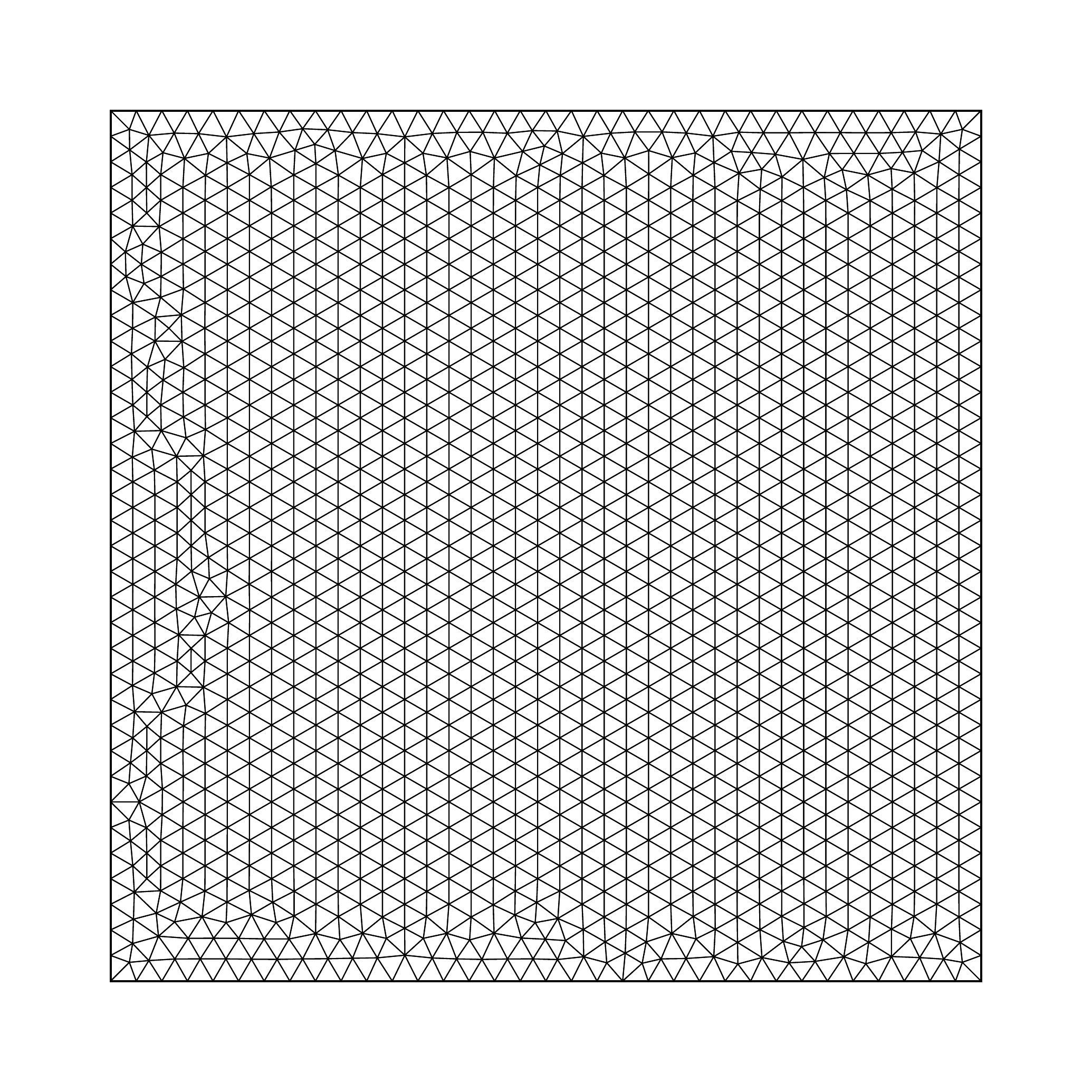} 	\end{subfigure}
        ~
    \begin{subfigure}[b]{0.32\textwidth}
        \includegraphics[width=\textwidth]{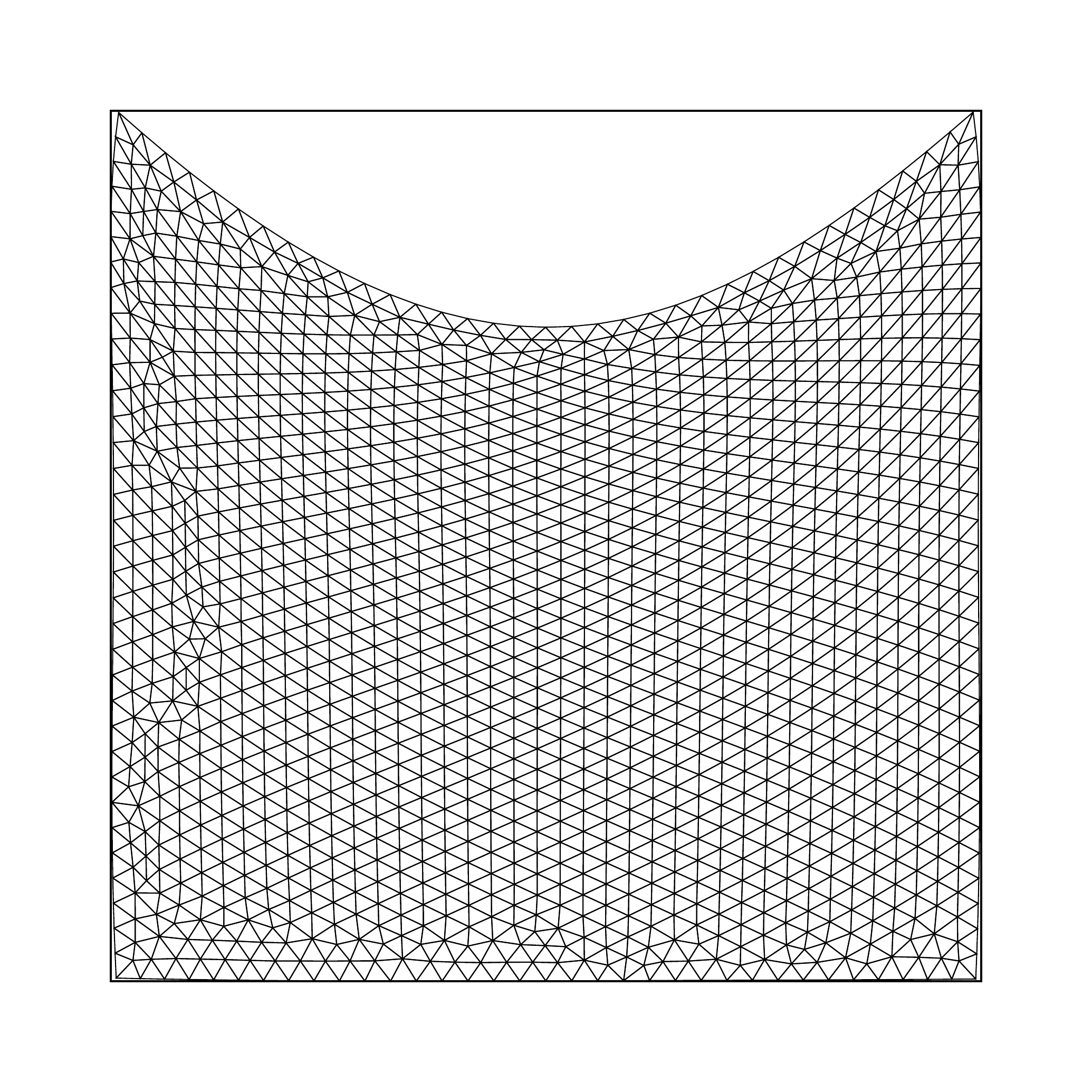}
    \end{subfigure}
        ~
    \begin{subfigure}[b]{0.32\textwidth}
        \includegraphics[width=\textwidth]{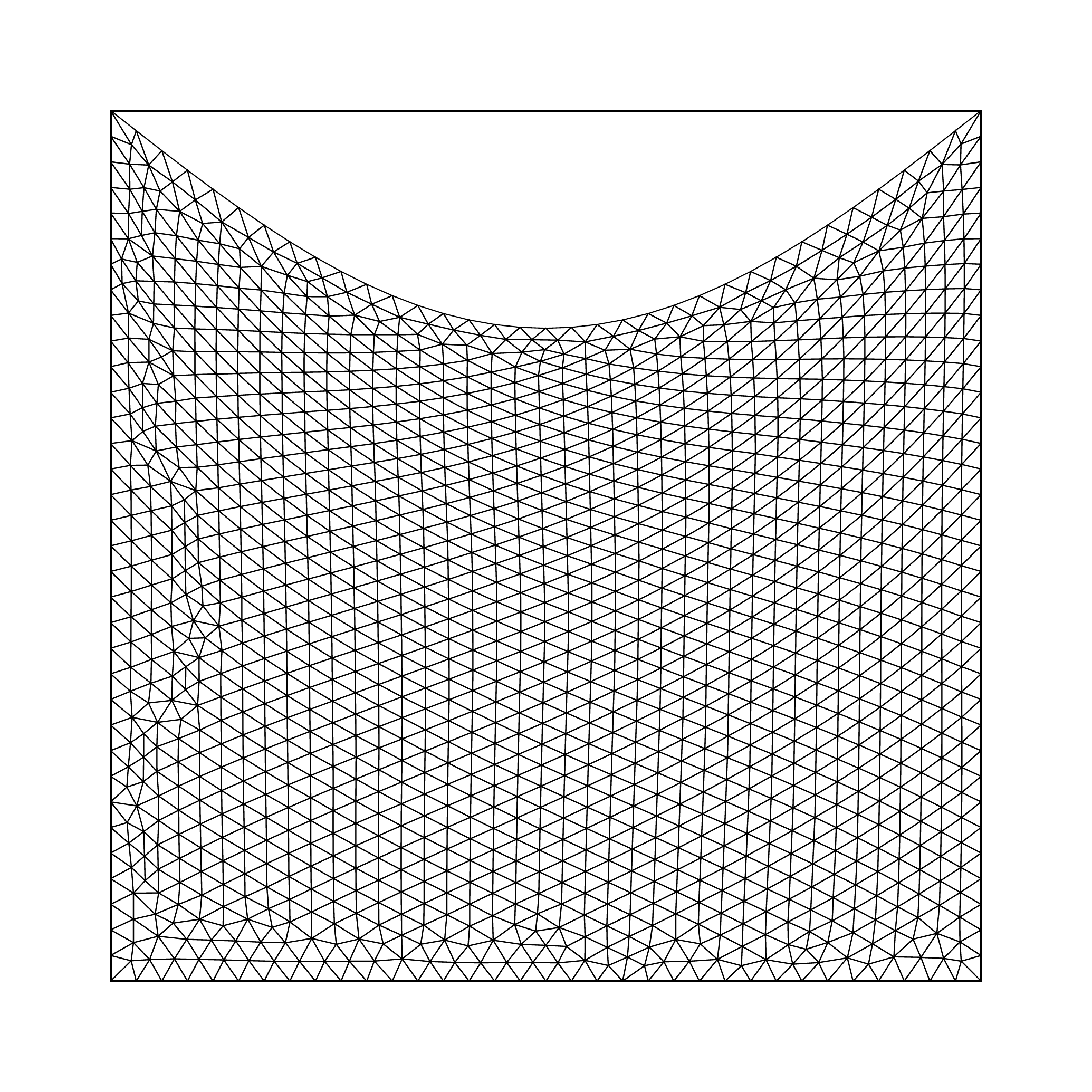}
    \end{subfigure}
        
	\end{center}
	\caption{Initial and deformed meshes of the deformed square case with its deformed top boundary. The unstructured mesh consists of 2744 triangular elements The first deformed figure shows the solution with classical PINN. The last figure represents the solution with exact boundary enforcement.}
	\label{fig.square_deformTop}
\end{figure}

In this test case, a square domain is deformed from its boundaries. The square domain is $x,\: y\in [0,1]\times[0,1]$ and the unstructured mesh consists of 2744 triangular elements. The initial mesh can be seen in Figure \ref{fig.square_deformTop}. We want to find a deformed mesh where the position of the top boundary becomes $\hat{y} = y - 0.25\sin(\pi x)$. On the top surface, we implement this condition as a Dirichlet boundary condition as well as $\hat{x} = x$. All the other boundaries have the same Dirichlet boundary condition as $\hat{y}=y$, and $\hat{x}=x$. The deformed mesh can be seen in Figure \ref{fig.square_deformTop}. The figure in the middle shows the results obtained by only using classical PINN. This shows the boundaries, especially the corners, are not in the exact position and are deformed in an undesired way. The figure on the right shows the solution after exact boundary enforcement. The boundary values are corrected with the exact positions with the proposed approach. The $L_2$ error on the boundary nodes is calculated as 0.031. For this test case, we increased the specific weight of the boundary loss of the composite loss function in Equation \ref{eq:pinnLoss}. Since the deformation of the boundary is higher than the deformation of the computational domain, the boundary weight is increased. The weight ratio of the boundary loss and the residual loss is set to 25 to capture the boundary values more precisely. 

\begin{figure}
	\begin{center}
      \begin{subfigure}[b]{0.45\textwidth}
			\includegraphics[width=\textwidth]{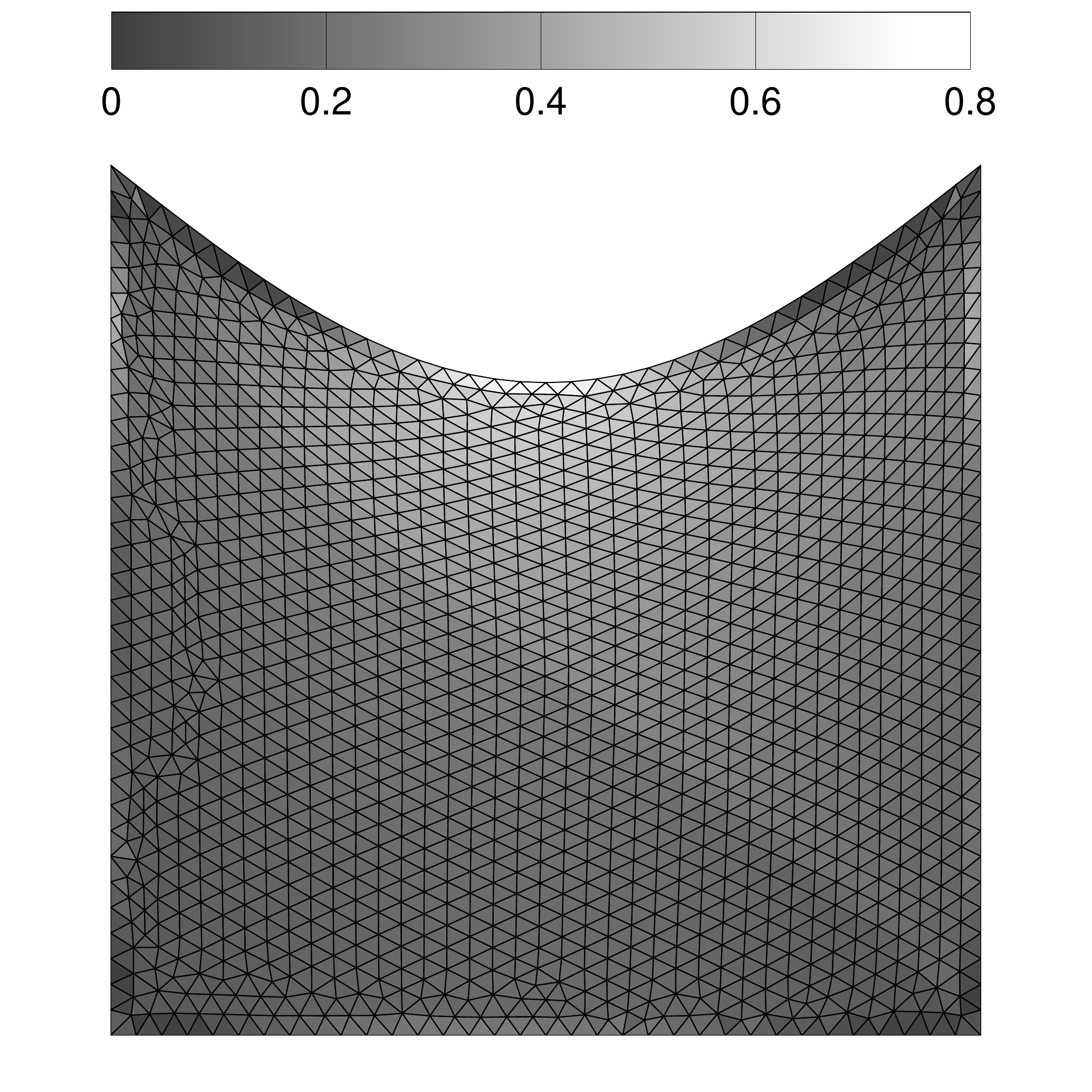} 
    \caption{Area change}
		\end{subfigure}
        ~
        \begin{subfigure}[b]{0.45\textwidth}
          \includegraphics[width=\textwidth]{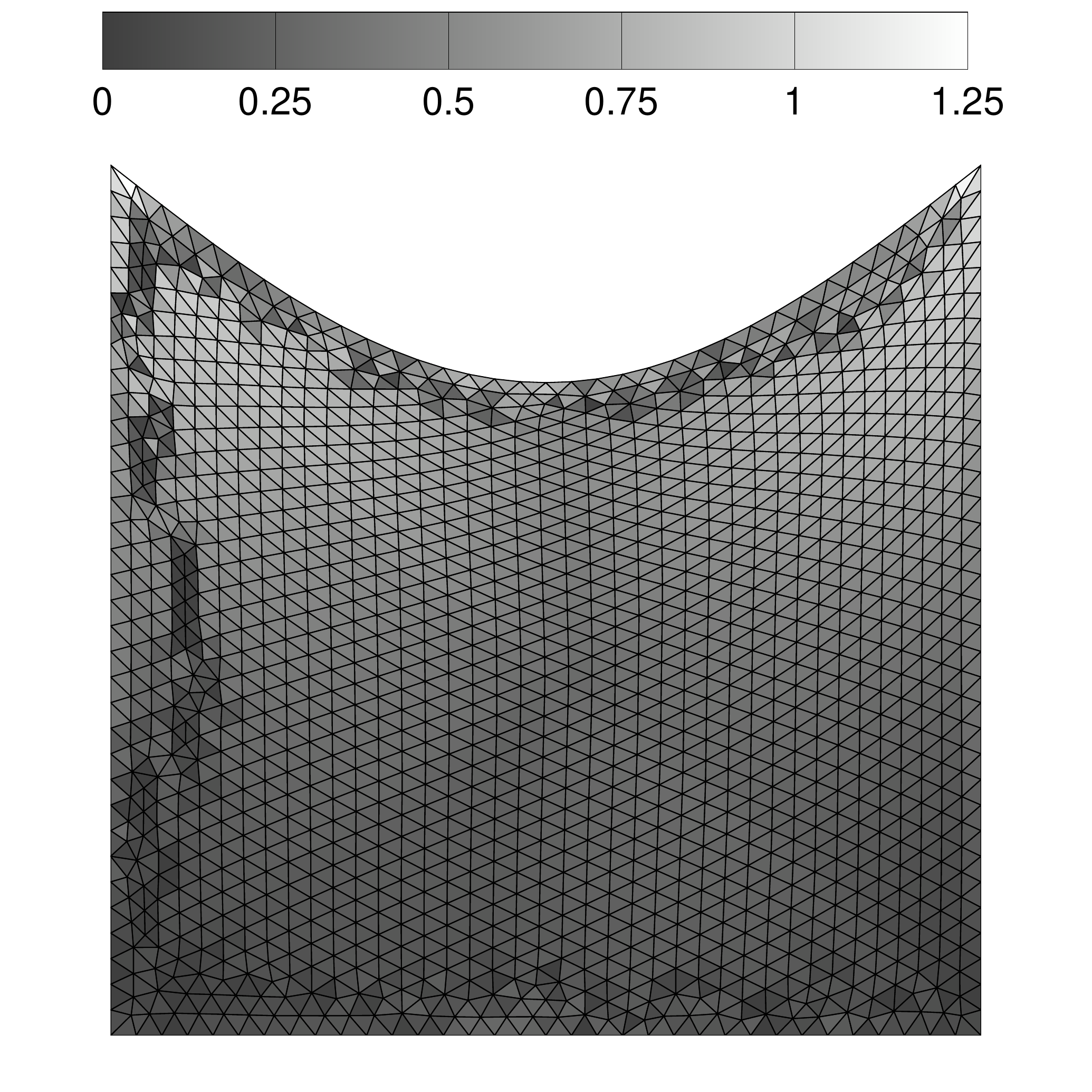}
          \caption{Shape change}
      \end{subfigure}
        ~
	\end{center}
	\caption{Element quality metrics of the square with deformed top boundary. The figure on the left shows the element area change and the figure on the right shows the element shape change with respect to the initial mesh elements.}
	\label{fig.squareMetric}
\end{figure}

The mesh quality measure of the deformed mesh based on the element area and shape changes can be seen in \ref{fig.squareMetric}. The top surface is deformed according to a sinusoidal function. The elements near the deformed boundary have the most change in size and shape as expected. Especially in the middle where the deformation is the largest, the elements are squeezed and get smaller. In the corners where the element vertices have two boundary conditions in each direction, the element area change is not significantly large. However, the shape of the corner elements changes more than the other elements on the boundary. These elements are bounded by the two boundaries and therefore the aspect ratios get larger. The deformation of the inner elements is relatively low, especially near the bottom boundary. The mesh deformation metrics get lower as the elements' position moves away from the deformed boundary. The global area and shape change metrics are calculated as  $|f_A^{\infty}| = 0.744$, $|f_{AR}^{\infty}| = 1.264$, respectively.

\begin{figure}
	\begin{center}
    \begin{subfigure}[b]{0.32\textwidth}
	\includegraphics[width=\textwidth]{figures/square/square_initMesh.pdf} 	\end{subfigure}
        ~
    \begin{subfigure}[b]{0.32\textwidth}
        \includegraphics[width=\textwidth]{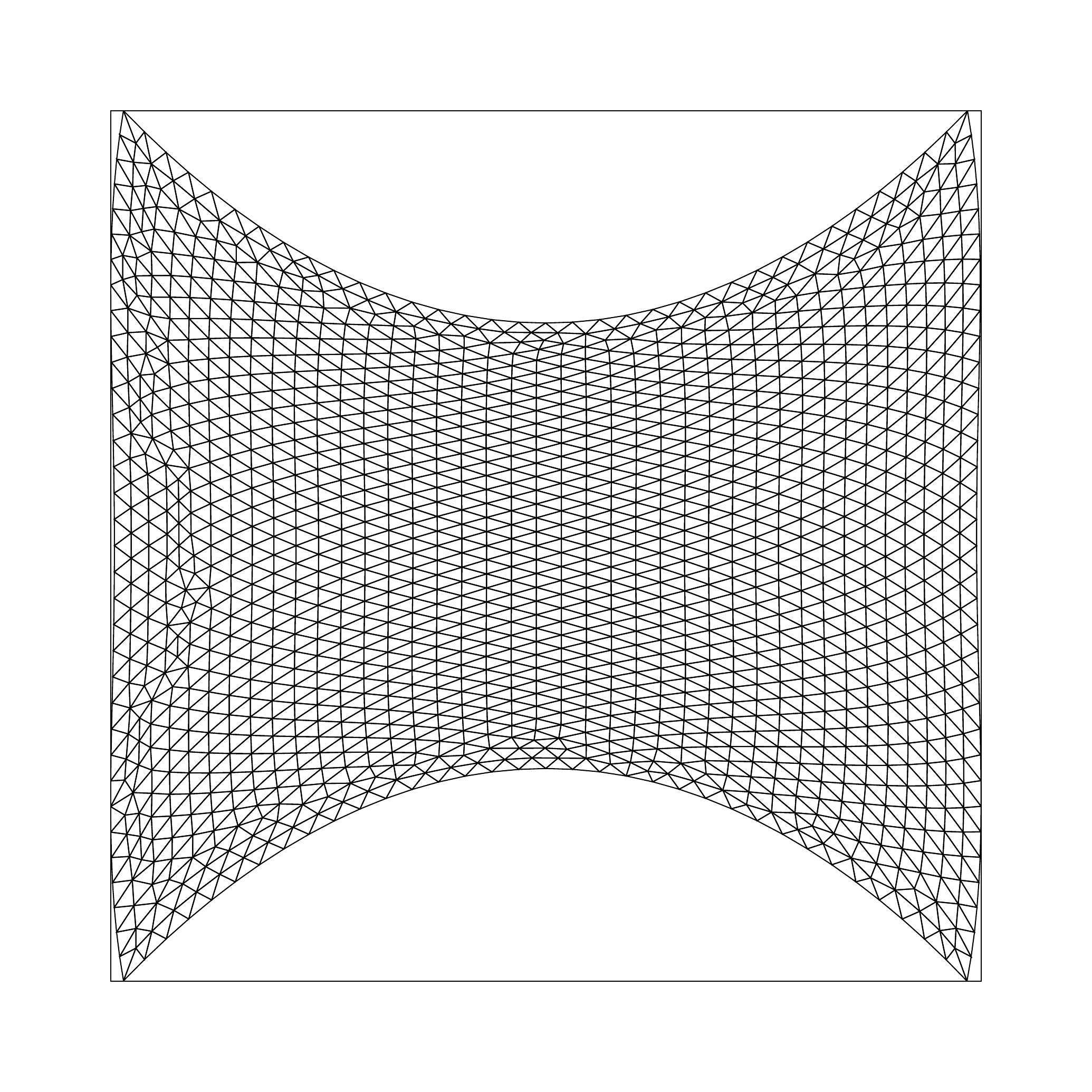}
    \end{subfigure}
        ~
    \begin{subfigure}[b]{0.32\textwidth}
        \includegraphics[width=\textwidth]{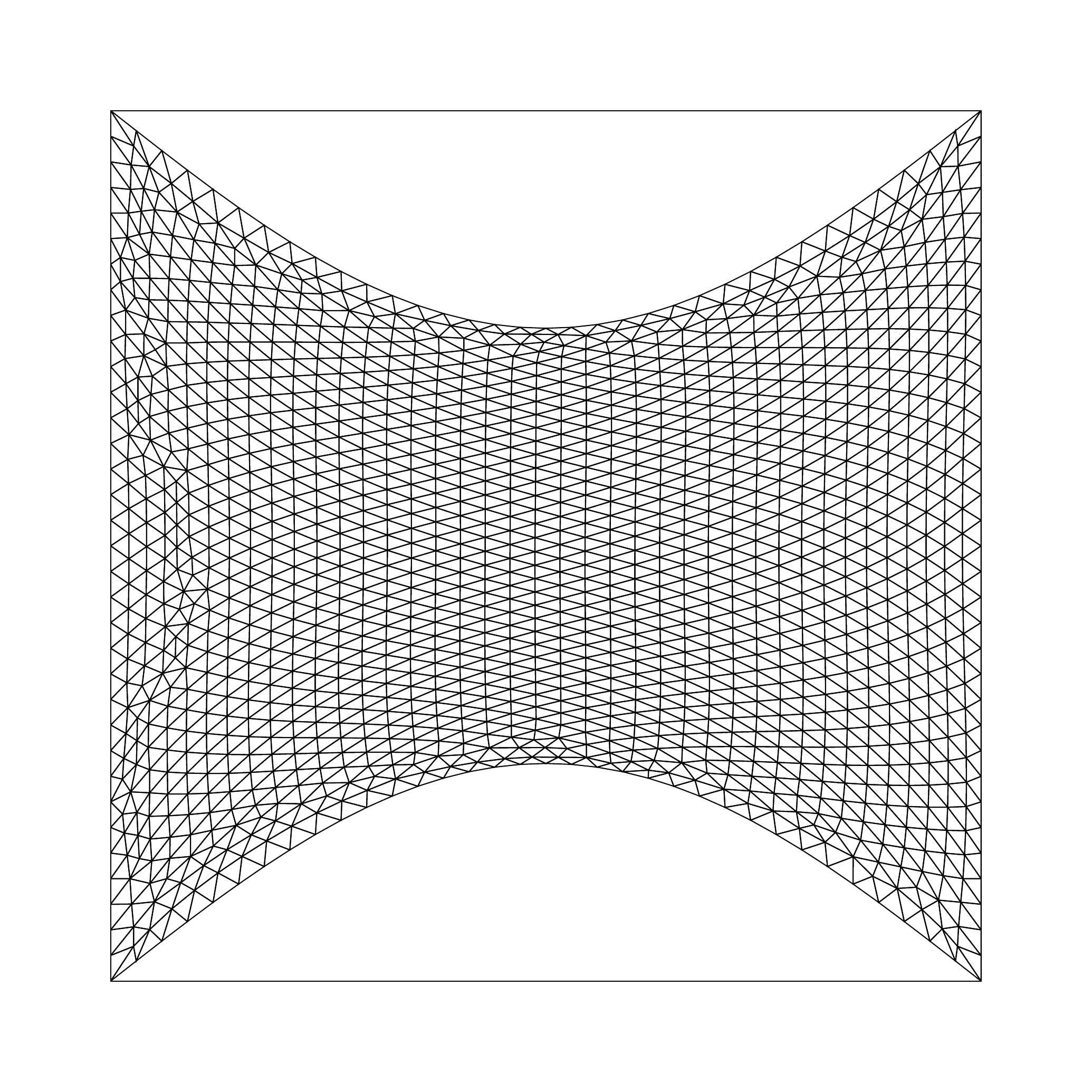}
    \end{subfigure}
        
	\end{center}
	\caption{Initial and deformed meshes of the deformed square case. The square is squeezed from its top and bottom boundary. The first deformed figure shows the solution with classical PINN. The last figure represents the solution with exact boundary enforcement.}
	\label{fig.square_squeeze}
\end{figure}

To see the capabilities of our approach we further deform the bottom boundary with its coordinates $\hat{y} = y + 0.25\sin(\pi x)$. The Dirichlet boundary conditions on the stationary boundaries are the same as the other, $\hat{x}=x$, $\hat{y}=y$. The same specific weight ratio for the loss function of the PINN formulation is used. The deformed configuration can be seen in Figure \ref{fig.square_squeeze}. The figure in the middle is the solution with the classical PINN approach. The vertices on the boundaries are not in exact positions. Especially on the corners, the classical PINN solution has difficulty satisfying the positions. The $L_2$ error of the boundary positions is calculated as 0.076 for this case. 

\begin{figure}
	\begin{center}
      \begin{subfigure}[b]{0.45\textwidth}
			\includegraphics[width=\textwidth]{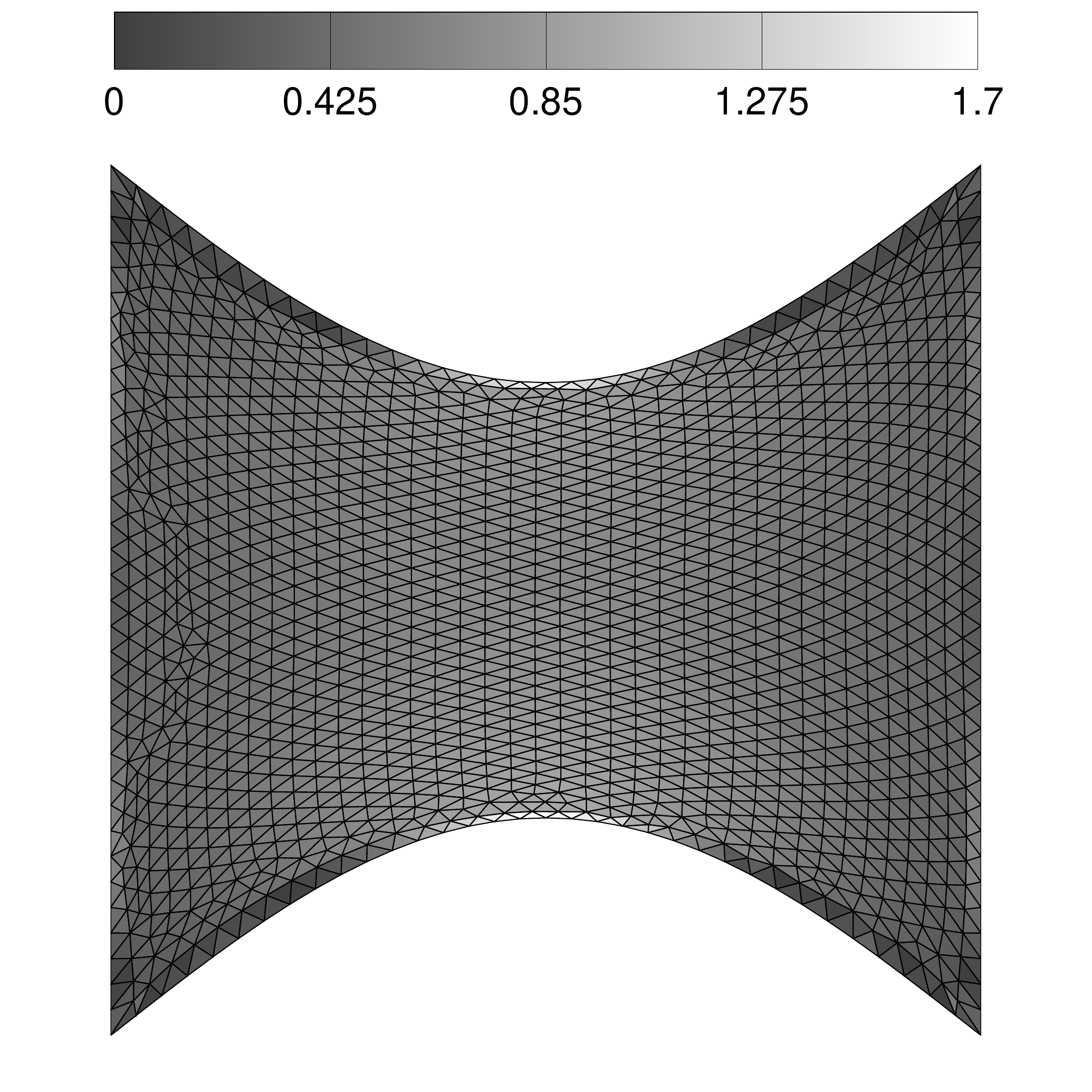} 
    \caption{Area change}
		\end{subfigure}
        ~
        \begin{subfigure}[b]{0.45\textwidth}
          \includegraphics[width=\textwidth]{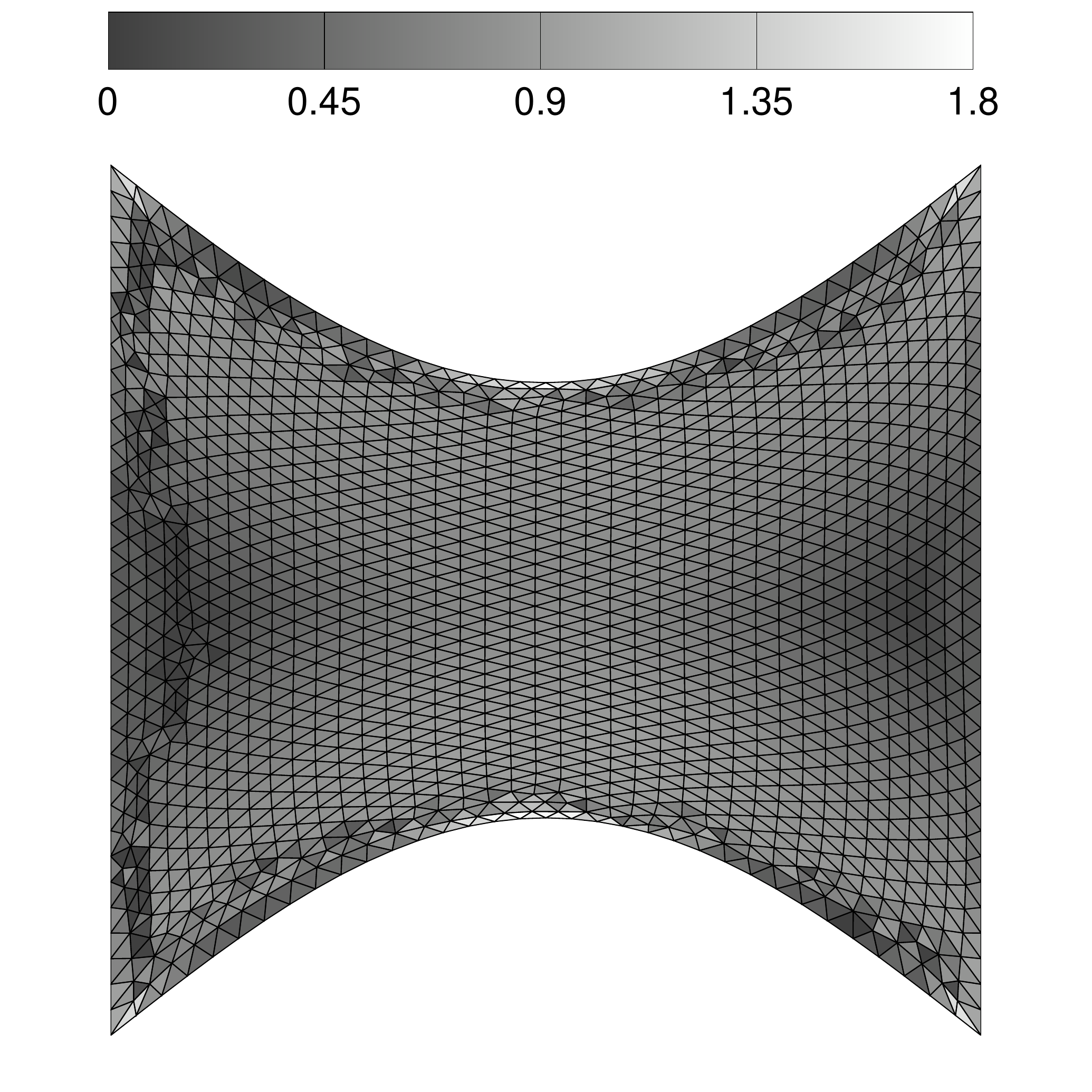}
          \caption{Shape change}
      \end{subfigure}
        ~
	\end{center}
	\caption{Element quality metrics of the square deformed from the top and bottom boundaries. The figure on the left shows the element area change and the figure on the right shows the element shape change with respect to the initial mesh elements.}
	\label{fig.squeezeMetric}
\end{figure}

The elementwise quality measures of this case can be seen in Figure \ref{fig.squeezeMetric}. the elements on the top and the bottom boundaries are deformed the most, the same as in the previous case. The elements in the middle collapsed more than the case before. The global area and shape change values are $|f_A^{\infty}| = 1.701$, $|f_{AR}^{\infty}| = 1.845$, respectively. The element shape and size change significantly as the deformation is increased.

\subsection{Translation and Rotation tests}
\begin{figure}[hbpt!]
	\begin{center}
      \begin{subfigure}[b]{0.32\textwidth}
	\includegraphics[width=\textwidth]{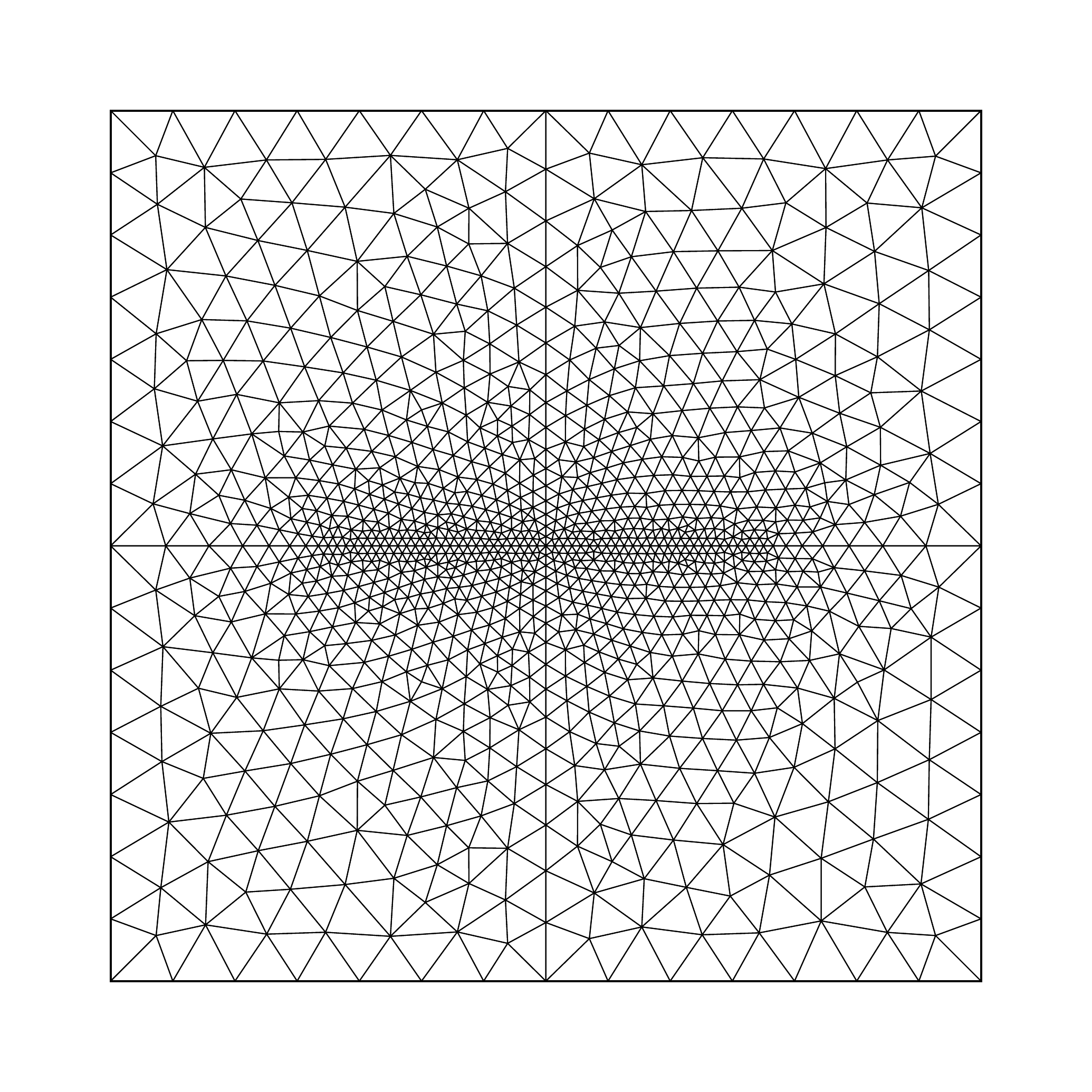} 
		\end{subfigure}
        ~
        \begin{subfigure}[b]{0.32\textwidth}
    \includegraphics[width=\textwidth]{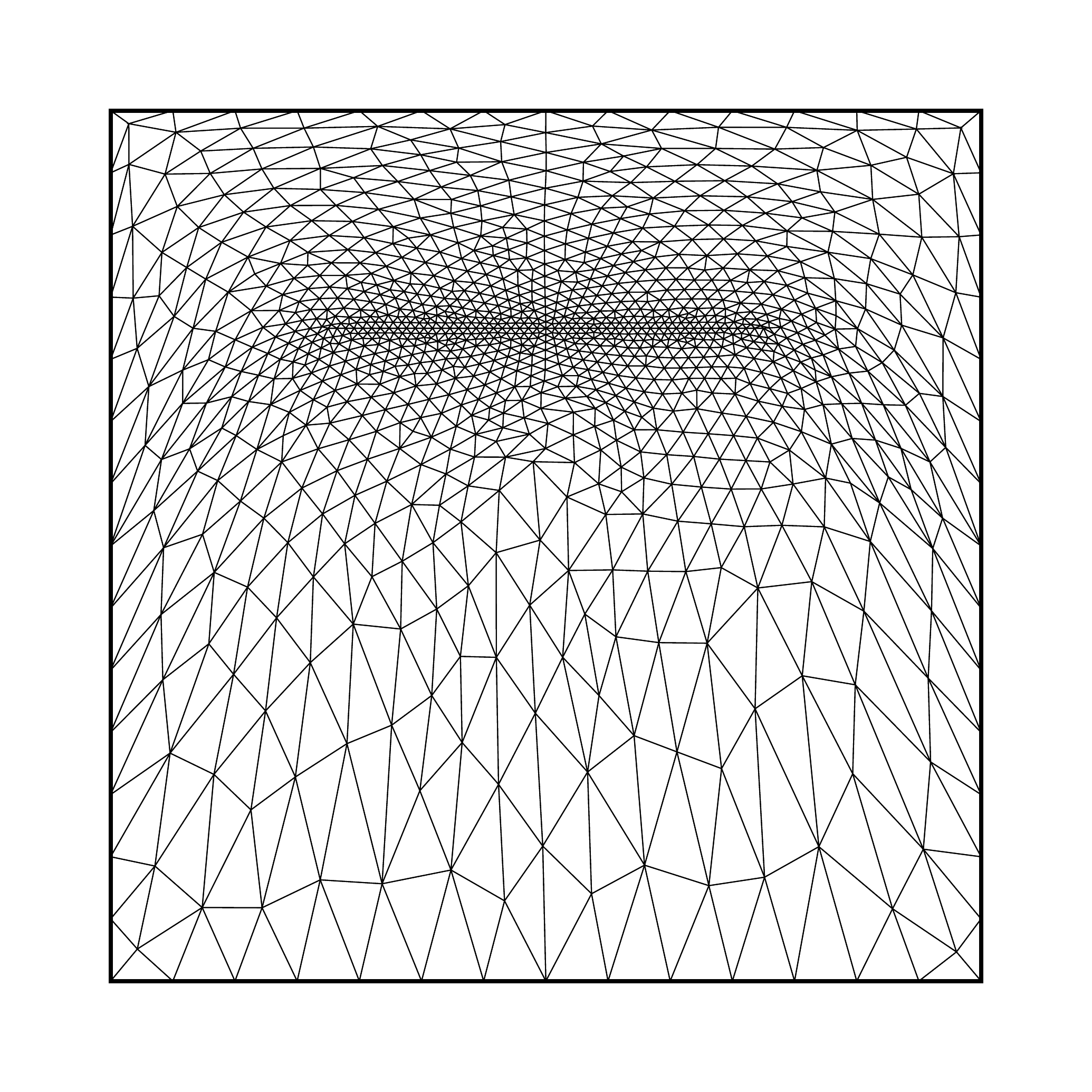}
      \end{subfigure}
        ~
        \begin{subfigure}[b]{0.32\textwidth}
   \includegraphics[width= \textwidth]{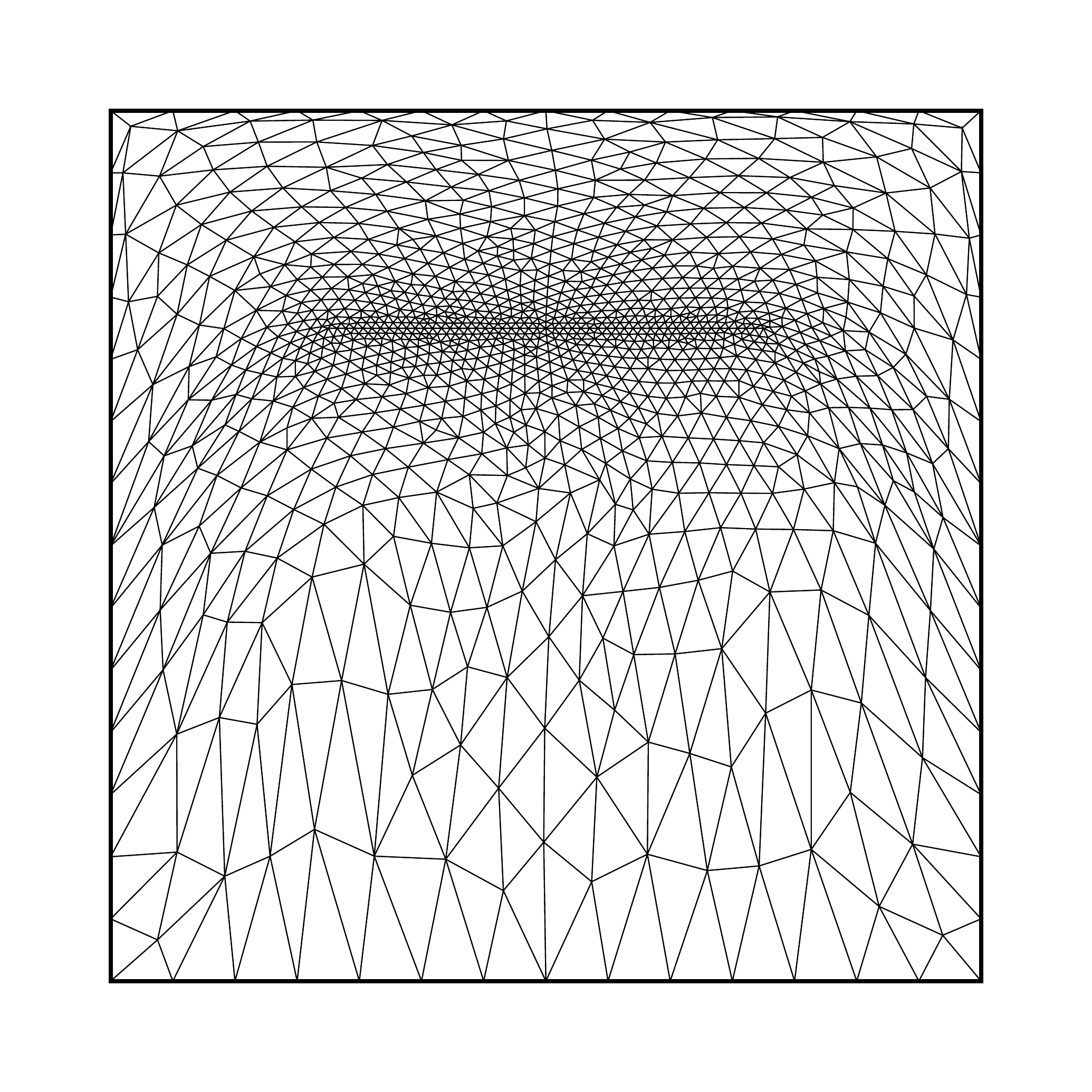} 
      \end{subfigure}
	\end{center}
	\caption{Initial mesh and deformed mesh after a total translation of $5$ units. The solution in the middle is performed in 10 steps while the solution on the right is performed in 5 steps.}
	\label{fig.translationTest}
\end{figure}

\begin{figure}[hbpt!]
	\begin{center}
      \begin{subfigure}[b]{0.32\textwidth}
	\includegraphics[width=\textwidth]{figures/Line/initial_mesh.pdf} 
		\end{subfigure}
        ~
        \begin{subfigure}[b]{0.32\textwidth}
    \includegraphics[width=\textwidth]{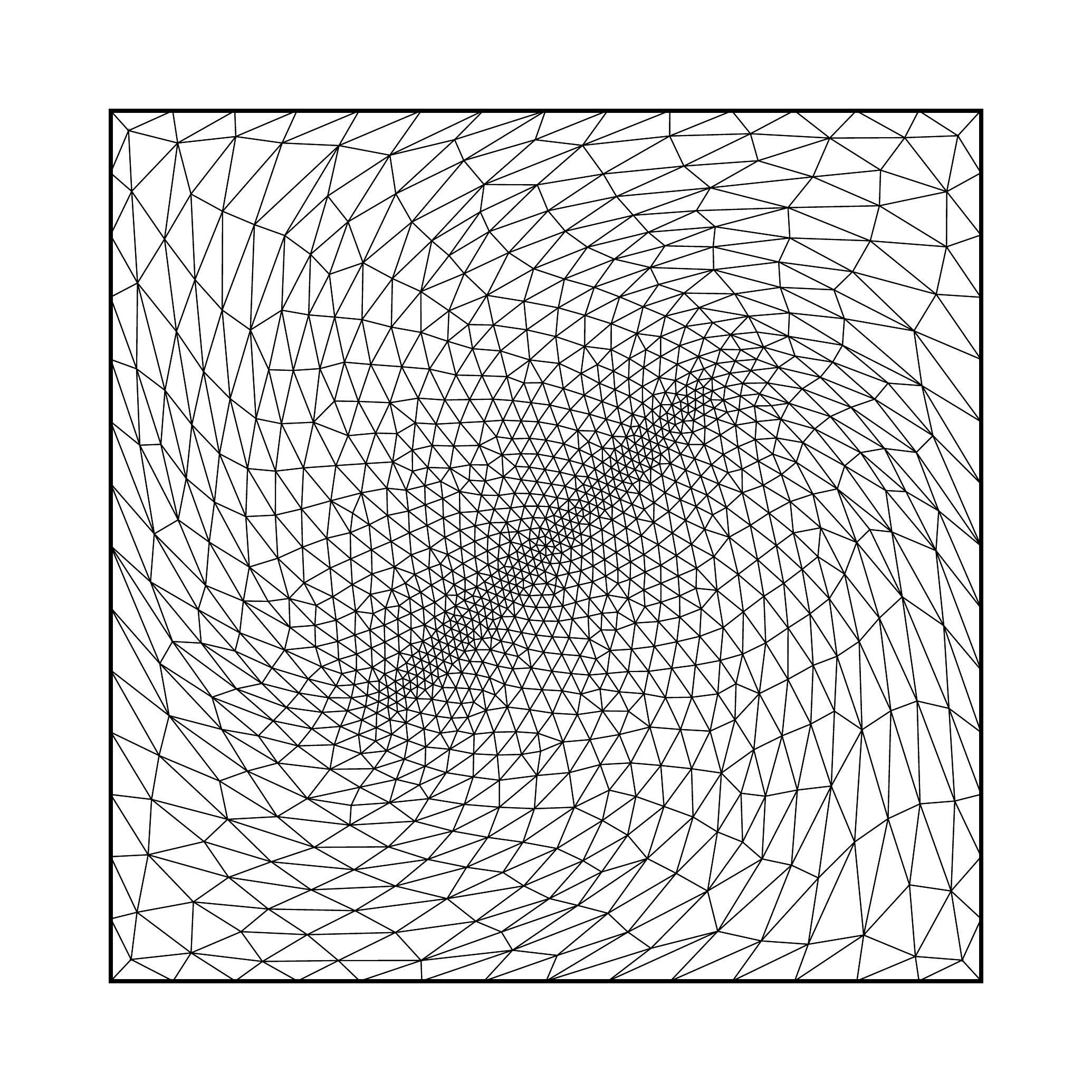}
      \end{subfigure}
        ~
        \begin{subfigure}[b]{0.32\textwidth}
   \includegraphics[width= \textwidth]{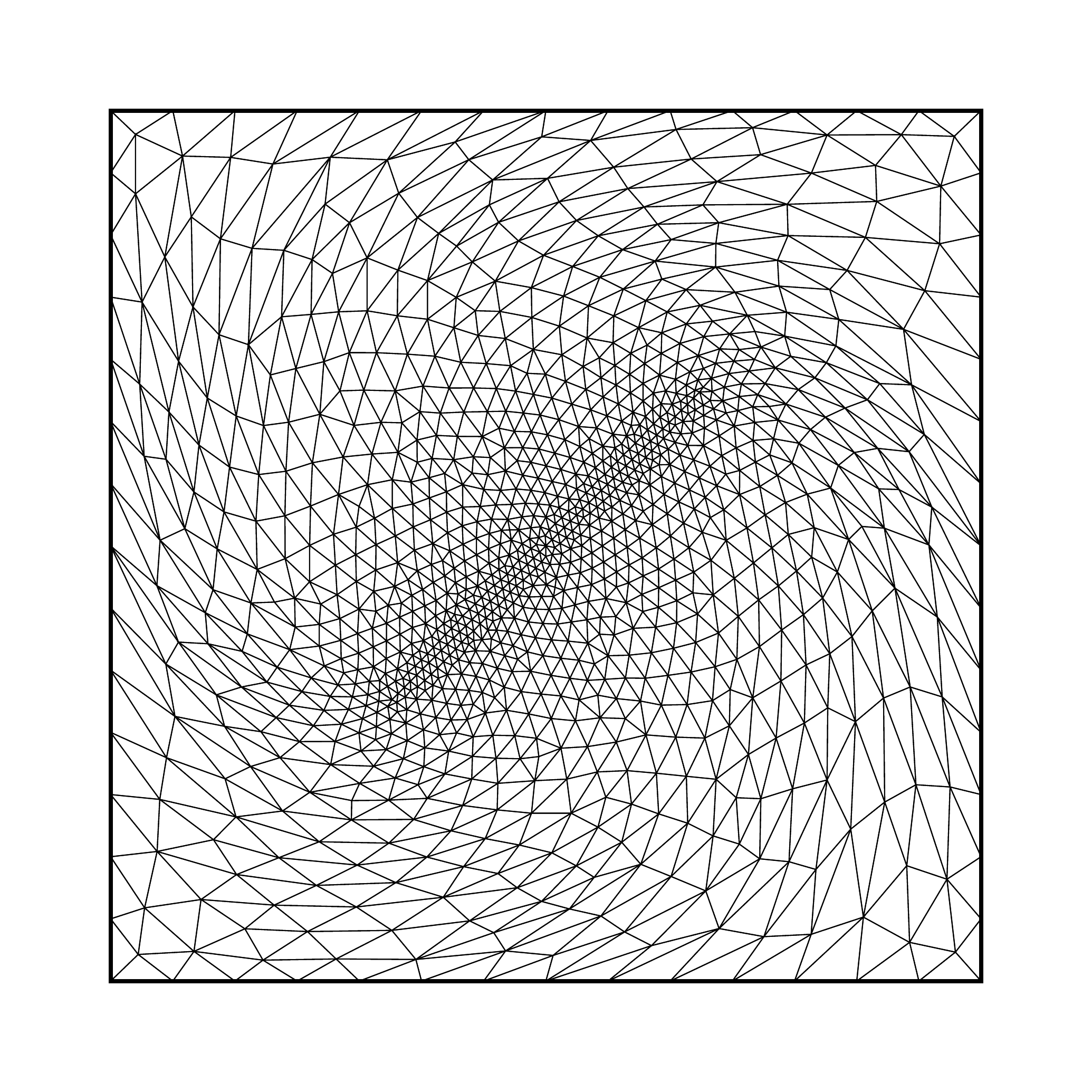} 
      \end{subfigure}
	\end{center}
	\caption{Initial mesh and deformed mesh after a total rotation of $0.25\pi$. The solution in the middle is performed in 10 steps while the solution on the right is performed in 5 steps.}
	\label{fig.rotationTest}
\end{figure}

To test the accuracy of our approach, translation and rotation tests in \cite{stein2003lineMesh} are performed. The original mesh can be seen in Figures \ref{fig.translationTest} and \ref{fig.rotationTest}. There is a line object located in $(-L, 0) \times (L,0)$ in a $(-2L, -2L)\times(2L, 2L)$ domain. A total of 2182 triangles are generated for the mesh.

\begin{figure}[hbpt!]
	\begin{center}
      \begin{subfigure}[b]{\textwidth}
	\includegraphics[width=\textwidth]{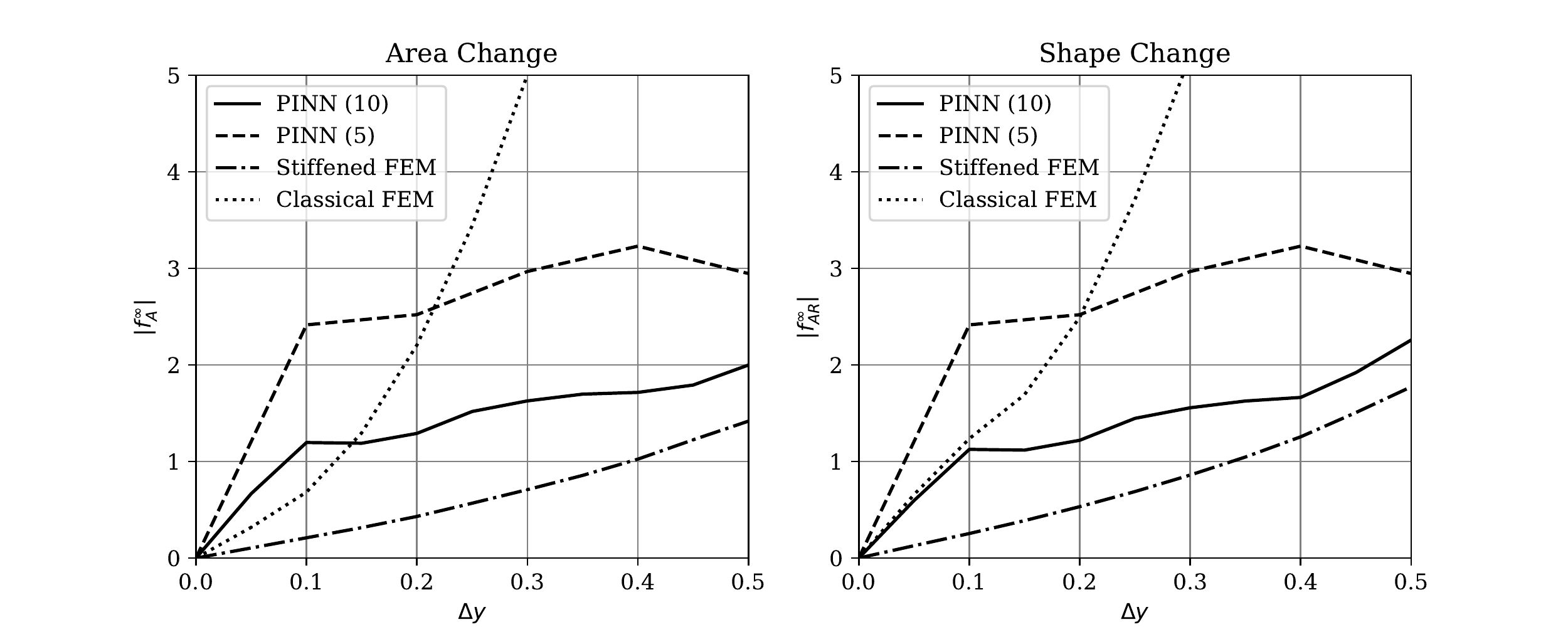} 
		\end{subfigure}
        ~
        \begin{subfigure}[b]{\textwidth}
    \includegraphics[width=\textwidth]{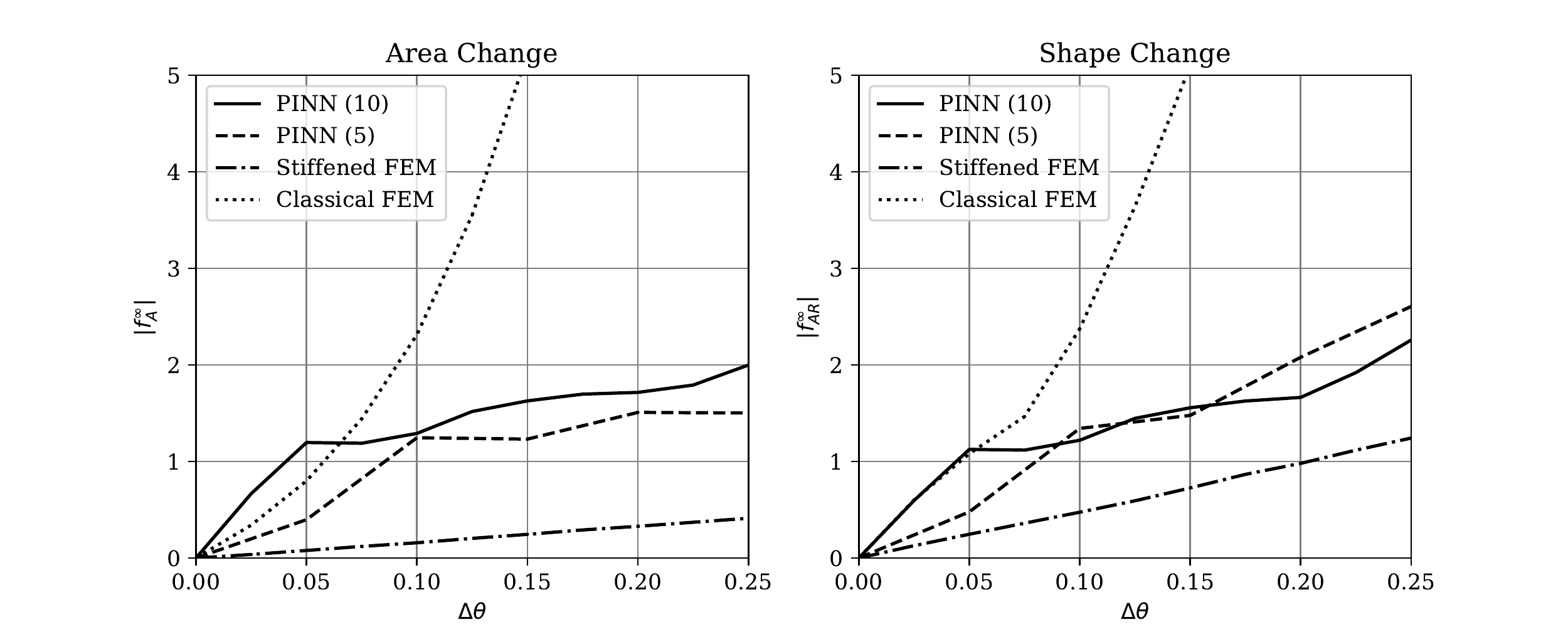}
      \end{subfigure}
	\end{center}
	\caption{Global area and shape change metrics of the translation and rotation tests compared with the FEM solution in \cite{stein2003lineMesh}. The first row shows the comparison of the translation test, while the second row shows the comparison of the rotation tests.}
	\label{fig.testPlots}
\end{figure}

For the translation tests, the object is moved $0.5L$ upwards. The movement is performed in 10 steps with $0.05L$ and in 5 steps with $0.1L$ movement upwards in two different training settings. The last step of the movement can be seen in Figure \ref{fig.translationTest}. In Figure \ref{fig.testPlots}, the PINN method is compared with the approach in \cite{stein2003lineMesh}. The area and shape change metrics of two PINN solutions are presented alongside the classical finite element solutions and solutions with Jacobian-based stiffening. The authors applied a stiffening power to prevent the deformation of the smaller elements. The stiffened approach represents the best value obtained in \cite{stein2003lineMesh} with different applied stiffening power. The two PINN solutions are representing the overall motion in 5 and 10 steps. The total number of steps is represented in parentheses in the figure. As seen in the first row of Figure \ref{fig.testPlots}, the PINN solutions are comparable with the FEM solutions with Jacobian-stiffening. As mentioned before, the PINN approach does not have any criteria to prevent mesh overlapping and sudden movements move the vertex nodes in an undesired way. Therefore, the quality of the deformed mesh improves as the number of steps increases. 

For the rotation tests, the object is rotated $0.25\pi$ counterclockwise. Again, to prevent overlapping of edges and collision of vertices, the movement is performed in steps with $0.025\pi$ and $0.05\pi$ counterclockwise movement in each step in two different training. The last step of the rotation can be seen in Figure \ref{fig.rotationTest}. The deformed mesh differs especially on the boundaries between different PINN solutions. The small elements near the moving boundary start to collapse in the PINN solution with 5 steps. As the number of steps increases, the mesh quality increases. The comparison of the rotation tests with the same finite element solution of the translation tests is presented in Figure \ref{fig.testPlots}. The PINN approach again lies between the classical solution and the solution with Jacobian-based stiffening. 

\begin{table}[htb!]
    \centering
    \caption{Global area and shape changes of translation tests. The solution is performed in 10 steps. The values are given in every step.}
    \begin{tabular}{ c | c c c c c c c c c c}
    \hline \hline
       $\Delta y$& 0.05 & 0.1 & 0.15 & 0.2 & 0.25 & 0.3 & 0.35 & 0.4 & 0.45 & 0.5 \\ \hline
        {$\lvert f_A\rvert_{\infty}$} & 0.667 & 1.196 & 1.189 & 1.291 & 1.518 & 1.627 & 1.697 & 1.715 & 1.791 & 1.998 \\ 
         {$\lvert f_{AR}\rvert_{\infty}$} & 0.596 & 1.125 & 1.118 & 1.220 & 1.447 & 1.556 & 1.626 & 1.663 & 1.921 & 2.258 \\ 
         \hline \hline
         
    \end{tabular}
    \label{tab:translationMetric}
\end{table}

\begin{table}[htbp!]
    \centering
    \caption{Global area and shape changes of rotation tests. The solution is performed in 10 steps. The values are given in every step.}
    \begin{tabular}{ c | c c c c c c c c c c}
    \hline \hline
      $\Delta\theta(\pi)$& 0.025 & 0.05 & 0.075 & 0.1 & 0.125 & 0.15 & 0.175 & 0.2 & 0.225 & 0.25 \\ \hline
       {$\lvert f_A\rvert_{\infty}$}  & 0.274 & 0.432 & 0.663 & 0.881 & 0.882 & 1.152 & 1.093 & 1.256 & 1.295 & 1.324 \\ 
         {$\lvert f_{AR}\rvert_{\infty}$} & 0.355 & 0.508 & 0.753 & 1.034 & 1.266 & 1.554 & 1.914 & 2.137 & 2.517 & 2.573 \\ 
          \hline \hline
         
    \end{tabular}
    \label{tab:rotationMetric}
\end{table}

In both tests, the global mesh quality metric presented in section \ref{ch:meshMotion} is used. The $\lvert f_A \rvert_{\infty}$ and $\lvert f_{AR} \rvert_{\infty}$ are calculated as the maximum area and shape change of the values in Equation \ref{eq:meshMetric} in every step. The area change and shape change values are presented in Tables \ref{tab:translationMetric} and \ref{tab:rotationMetric}, for the translation and rotation tests, respectively.


\subsection{Flexible Beam}
\begin{figure}[hbtp!]
	\begin{center}
      \begin{subfigure}[b]{0.45\textwidth}
	\includegraphics[width=\textwidth]{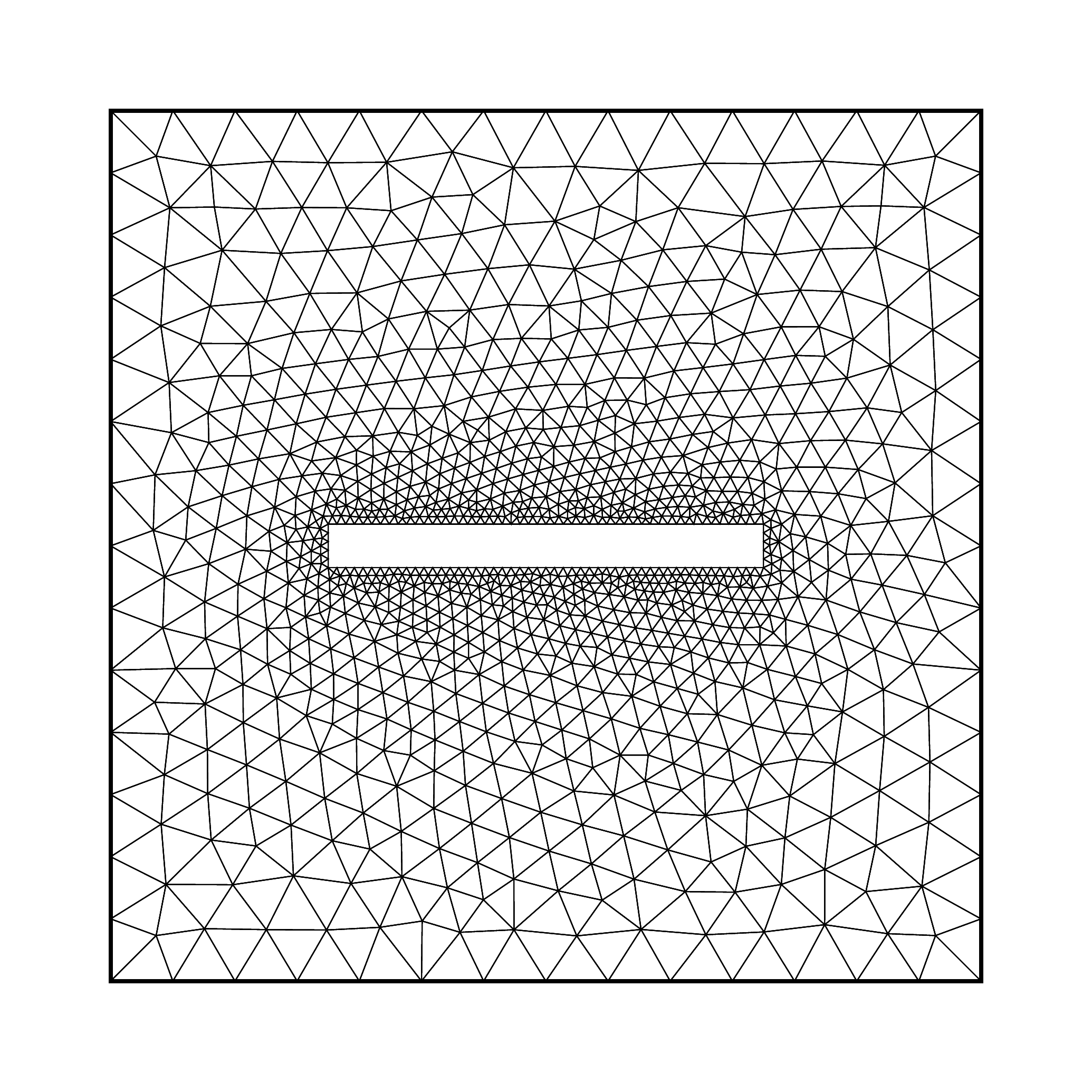} 
		\end{subfigure}
        ~
        ~
	\end{center}
	\caption{Initial mesh of the flexible beam test case with 2098 triangular elements. The elements are concentrated on the moving boundary to track the deformation in a precise way.}
	\label{fig.wingInitMesh}
\end{figure}

This test case consists of a mesh movement due to a motion of a flexible beam adapted from the problem in \cite{shamanskiy2021mesh}. The beam is fixed on its left end and sits in the center of the domain. Domain dimensions are $(-10,10)\times(-10,10)$ and the structure's position is $(-5,5)\times(-0.5,0.5)$ The deformation is based on a sinusoidal function $\sin (\frac{\pi}{2}\frac{x}{L})$ with varying amplitude. The initial mesh can be seen in Figure \ref{fig.wingInitMesh}. This unstructured mesh consists of 2098 triangular elements. The right end of the structure first moves to 4 units upwards, then 8 units downwards, following a 4-unit upward motion to return to its initial state. The movements are performed in steps with 2-unit motions, upwards or downwards. 

\begin{figure}[hbtp!]
	\begin{center}
      \begin{subfigure}[b]{0.45\textwidth}
			\includegraphics[width=\textwidth]{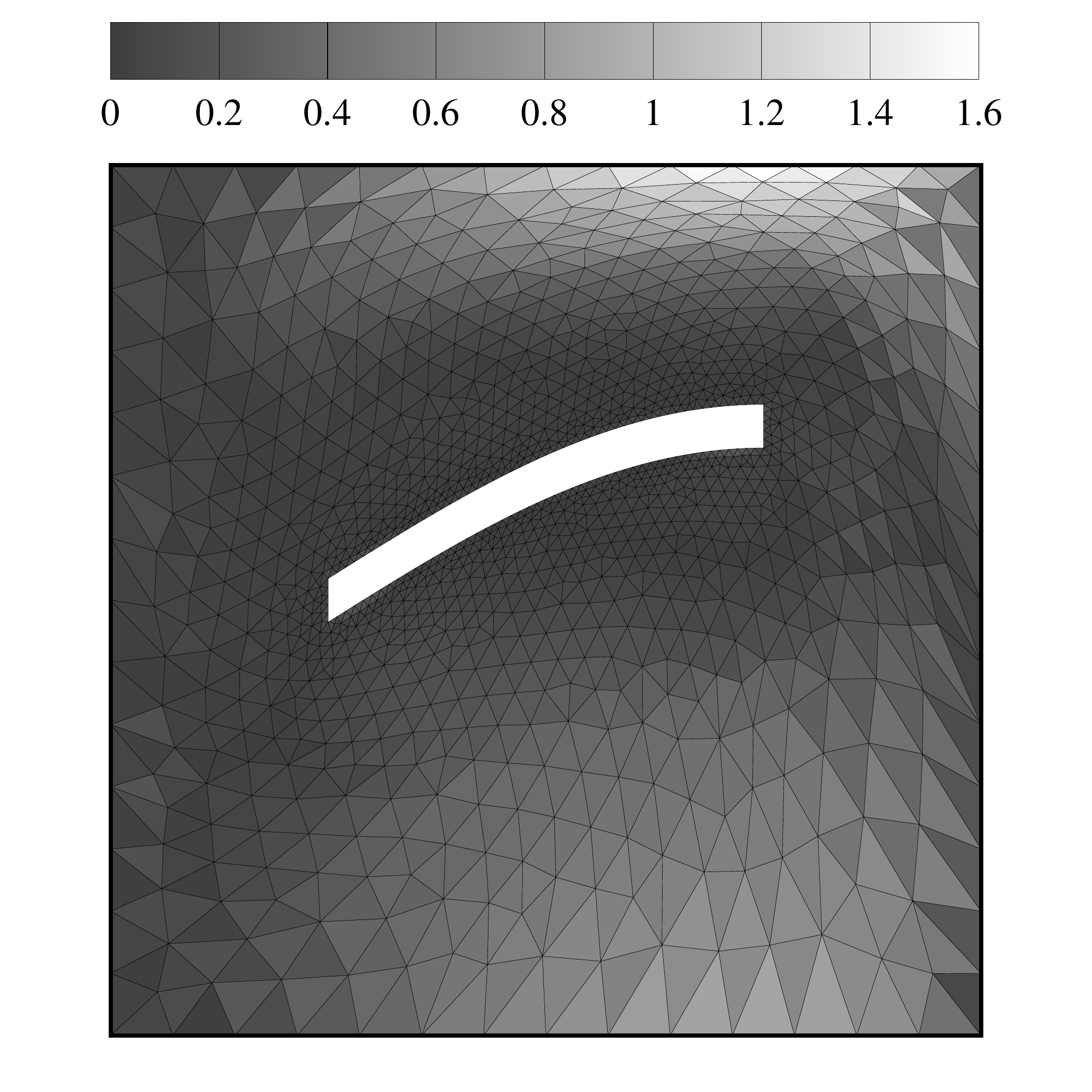} 
    \caption{Area change}
		\end{subfigure}
        ~
        \begin{subfigure}[b]{0.45\textwidth}
          \includegraphics[width=\textwidth]{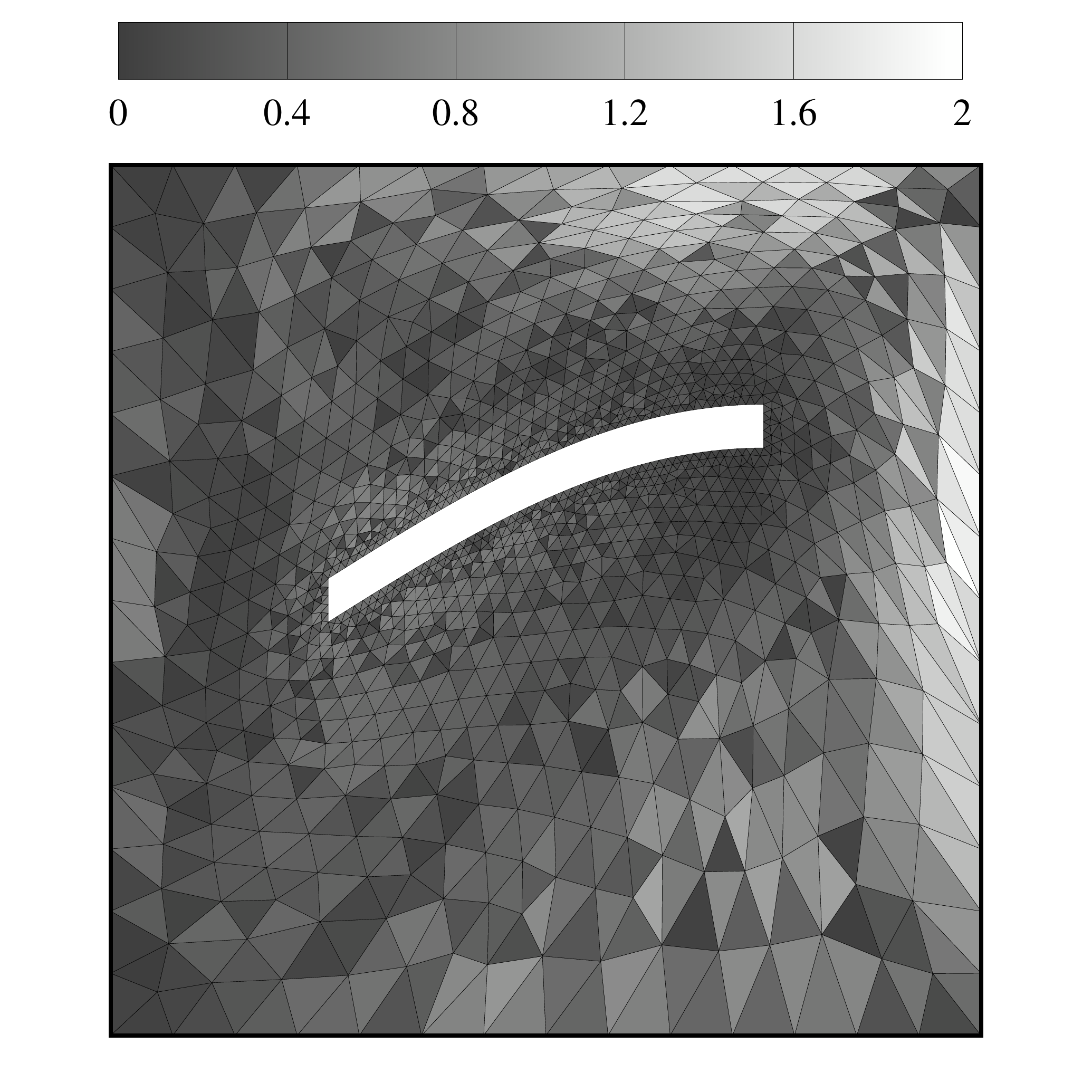}
          \caption{Shape change}
      \end{subfigure}
        ~
	\end{center}
	\caption{Element quality metrics when the structure tip moves to $y=4$.}
	\label{fig.wing2}
\end{figure}

In Figure \ref{fig.wing2}, the deformed mesh after two steps of movement is presented with the mesh quality presented with the global area and shape change metric. Using exact boundary enforcement gives the true boundary position and therefore fixes the vertices on the boundaries. Therefore, on the outer boundaries, elements are stretched and squeezed more than the inner elements. Especially elements near the tip of the moving boundary have the most area and shape changes. 

\begin{figure}[hbtp!]
	\begin{center}
      \begin{subfigure}[b]{0.45\textwidth}
			\includegraphics[width=\textwidth]{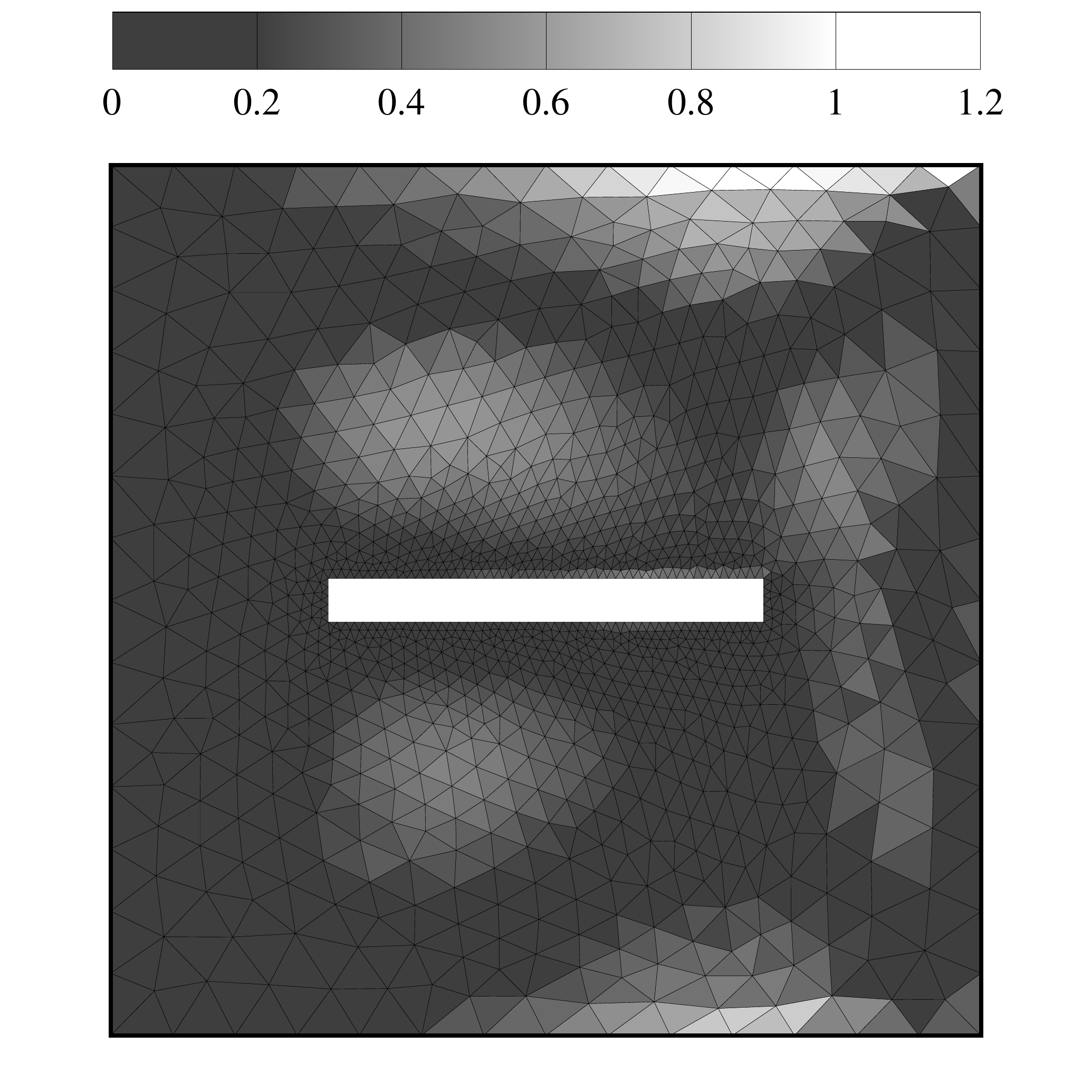} 
    \caption{Area change}
		\end{subfigure}
        ~
        \begin{subfigure}[b]{0.45\textwidth}
          \includegraphics[width=\textwidth]{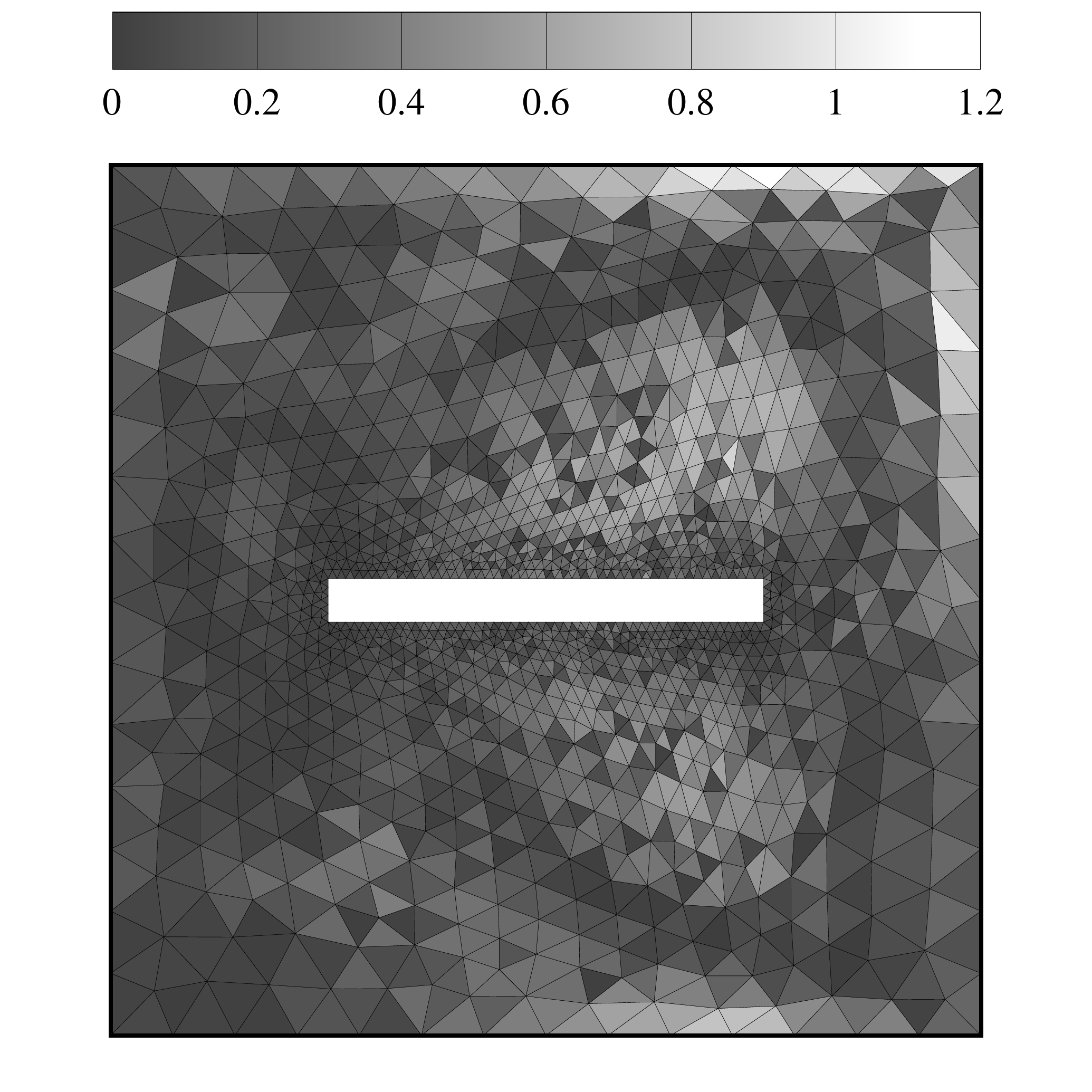}
          \caption{Shape change}
      \end{subfigure}
        ~
	\end{center}
	\caption{Element quality metrics when the structure returns its original position.}
	\label{fig.wing8}
\end{figure}

In figure \ref{fig.wing8}, the mesh after one cycle of motion is presented. The structure returns to its original place after eight iterations. By looking at the area change, the sinusoidal motion of the structure can be observed. The most deformed elements are located at the top and bottom boundaries and near the moving tip of the structure. These elements are squeezed first and cannot recover themselves after the relative stretching. 

\section{Conclusion and Future Work}
In this work, we solved mesh deformation problems with physics-informed neural networks. The selected method uses the linear elastic model since PINN can give accurate results for solving this type of PDE. We note that vertex nodes are moved according to boundary movement. Moreover, exact boundary values are enforced to satisfy the Dirichlet boundary conditions exactly. We test this approach with translation and rotation tests and compared it with finite element solutions. We showed that the PINN solution is comparable with the FEM solutions. The deformation is performed in numerous steps instead of a sudden movement. This prevents vertex collision and edge overlapping. We showed that as the number of steps is increased, the deformed mesh quality gets higher. For a greater mesh quality, the number of steps can be increased.

The mesh movement method in this paper only includes linear elastic equations, although it can be extended to other techniques. Other commonly used methods such as the Laplacian or biharmonic equations are also applicable to PINN formulation. Our future work aims to use other methods that prevent mesh overlapping in the training of the PINN. The network parameters and formulation can be extended in a way that vertex collisions and edge overlapping is prevented.

\bibliographystyle{ieeetr}
\bibliography{main}

\end{document}